\documentclass[preprints,article,accept,moreauthors,pdftex,10pt,a4paper]{Definitions/mdpi}

\firstpage{1}
\pubvolume{xx}
\issuenum{1}
\articlenumber{5}
\pubyear{2018}
\copyrightyear{2018}
\history{Received: date; Accepted: date; Published: date}
\usepackage{amsmath}
\usepackage{color,soul}
\Title{Aggregating Large-Scale Generalized Energy Storages to Participate in Energy Market and Regulation Market}

\Author{Yao Yao $^{1}$, Peichao Zhang $^{1,}$*, Sijie Chen$^{1}$}

\AuthorNames{Yao Yao and Peichao Zhang and Sijie Chen}

\address[1]{%
$^{1}$ \quad Key Laboratory of Control of Power Transmission and Conversion of Ministry of Education, Department of Electrical Engineering, Shanghai Jiao Tong University, Minhang District, Shanghai 200240, China}

\corres{Correspondence: pczhang@sjtu.edu.cn}

\abstract{ This paper proposes a concept of generalized energy storage (GES) to facilitate the integration of large-scale heterogeneous flexible resources with electric/thermal energy storage capacity to participate in multiple markets. First, a generalized state variable referred to as degree of satisfaction (DoS) is defined, and dynamic models with a unified form are derived for different types of GESs. Second,  a real-time market-based coordination framework is proposed to facilitate control, and ensure user privacy and device security. Demand curves of different GESs are then developed based on DoS to express their demand urgencies as well as flexibilities. Furthermore, a low-dimensional aggregate dynamic model of a GES cluster is derived thanks to the DoS-equality control feature provided by the design of demand curve. At last, an optimization model for a large-scale GESs to participate in both the energy market and regulation market is established based on the aggregate model. Simulations results demonstrate that the optimization algorithm could effectively reduce the total cost of an aggregator. Additionally, the proposed coordination method has high tracking accuracy and could well satisfy users' diversified power demand.}


\begin{document}

\section{Introduction}
The increasing integration of distributed energy resources (DERs) poses threat on stability and reliability of power system operation. With the development of smart grid technologies, the control of flexible loads, e.g., electric energy storage (EES), thermostatically control load (TCL) and electric vehicle (EV), has become a promising research field to address the problem brought by DERs \cite{Hao2017Optimal,Mingshen2018Load}. Due to the great variety, large scale, wide distribution and small individual capacity of flexible loads, load aggregators (LAs) are required to aggregate and dispatch such loads and provide flexible services to power grid. The coordination strategies have been widely studied. For example, a three-step EV charging algorithm is presented in \cite{Vandael2013A} to minimize charging cost. In \cite{Hao2014Aggregate}, a priority-stack-based method is put forward for fixed-frequency air-conditioner (FFA) fleet to track automatic generation control signal. Literature \cite{Song2017Thermal} establishes physical model of inverter air-conditioners (IVAs) and proposes a hierarchical control framework. A unified state model for EVs and TCLs is developed in \cite{Mingshen2018Load}. In \cite{Hao2017Optimal}, a generalized battery model is presented to describe the operational characteristics of commercial/residential building HVAC and energy storage system, based on which optimal control methods are proposed to provide various services.

However, most of these methods have the following disadvantages. First, most approaches can only be applied to a certain type of load \cite{Vandael2013A,Hao2014Aggregate,Song2017Thermal}. An LA should provide different interfaces to integrate different types of resources, which makes it difficult for the LA to fully utilize flexibility of various flexible loads and increase the control cost. Second, many literatures use centralized control method based on direct load control \cite{Hao2014Aggregate,Mingshen2018Load,Hao2017Optimal}. An LA should collect detailed parameters of all controlled loads and specify the response power of each load, which has heavy computational burden and communication traffic, and may lead to privacy issues and device security problems. Such methods may not be suitable for coordination of large-scale flexible loads. Some papers develop unified models to coordinate different types of resources \cite{Mingshen2018Load,Hao2017Optimal}, while they depend on centralized control as they mainly focus on individual behaviour rather than aggregate performance.

The objective of LAs is to provide flexible services to power grid. The method in \cite{Hao2014Aggregate} could provide reliable regulation services. Literature \cite{Mingshen2018Load} demonstrates its effectiveness in power fluctuation smoothing. In \cite{Vandael2013A,Song2017Thermal}, LAs participate in optimal dispatch and achieve benefits from energy markets. However, the above researches only consider a single market. It is pointed in \cite{Namor2018Control} that since different services have different requirements, participating in multiple markets helps to better utilize the control flexibility of resources and obtain more benefits. To stack multiple services, energy storage systems are controlled in \cite{Dubey2017Framework} to simultaneously participate in N-1 contingency requirement, voltage management and frequency regulation. An optimal control method is presented in \cite{Wu2015Energy} for energy storages to provide grid services including energy arbitrage, balancing service, capacity value and distribution system deferral, and outage mitigation. \cite{Cheng2018Co} and \cite{Anderson2017Co} introduce a Markov decision process model and a stochastic control method respectively to co-optimize battery storage for energy arbitrage and frequency regulation. However, these papers only consider energy storages. Although literature \cite{Hao2017Optimal} considers coordination of various flexible loads to provide multiple grid services, it has to solve some centralized optimization problems, which may lead to a high computational cost.

To address the above problems, a unified modelling method and coordination strategy for generalized energy storages (GESs) is proposed. It has unified information interface and control method, and requires a relatively low communication and computation cost, which is helpful for an LA to conduct coordination control over large-scale GESs. The contributions of our work are threefold:
\begin{enumerate}
\item Dynamic models with a unified form are developed for heterogenous GESs based on a generalized state variable referred to as degree of satisfaction (DoS).
\item A real-time coordination framework based on the market equilibrium mechanism is presented to allocate aggregate power to individual GESs. General demand curves are constructed under the framework to achieve equal degree of satisfaction and meet users' diversified requirements.
\item A low-dimensional aggregate dynamic model for a GES cluster is derived and used in an optimization model for an LA to participate in both the energy market and regulation market.
\end{enumerate}

The rest of this paper is organized as follows. Section \ref{sec:dynamic_model} introduces the dynamic model of different GESs. In section \ref{sec:coordination_method}, the market-based real-time coordination framework is proposed and demand curve construction methods for GESs are presented. Section \ref{sec:aggregate_model} derives the aggregate dynamic model for a GES cluster. In section \ref{sec:application}, the optimization problem that considers both energy and regulation markets is given. Section \ref{sec:simulation} shows the simulation results which demonstrate the effectiveness of our method. Finally, section \ref{sec:conclusion} summarizes our contribution and future work.

\section{Dynamic Models for GESs}
\label{sec:dynamic_model}

This paper focuses on four typical types of GESs, i.e., EES, EV, inverter air-conditioner (IVA) and fixed-frequency air-conditioner (FFA) since they account for a large share of flexible resources in the demand side. These resources are called GES for that they all can store energy, i.e., electric energy or cold/thermal energy, thus their power consumption could be adjusted without affecting the user satisfaction. Meanwhile, they have similar dynamic characteristics, which makes it possible to establish a unified physical model for them.  For example, there have been a few literatures that develop battery modelling method for the thermal energy storage loads. However, most researches only focus on FFAs\cite{Hao2013A,Mathieu2015Arbitraging,Sanandaji2014Improved}. A battery-type reduced-order model for building HVAC system is proposed in \cite{Hughes2016Identification}, but it only considers the dynamics of individual load.  Literature \cite{Song2017Thermal} studies the battery modelling for individual IVA and aggregated IVA cluster, while the aggregating strategy would lead to the result that the IVA cluster's regulation ability could not be fully utilized. This paper will put forward a unified dynamic model for different types of GESs, and study both individual model (introduced in this section) and aggregate model (detailed in section.\ref{sec:aggregate_model}).

\subsection{Degree of Satisfaction, DoS}
Before establishing the dynamic models, a dimensionless state variable referred to as degree of satisfaction (DoS) is defined for GESs, with the following purposes:
\begin{enumerate}
\item It could be used to measure the user satisfaction. The range of DoS is set to [-1,1], and the closer DoS is to 0, the higher user satisfaction is.
\item DoS could reflect a GES's state of energy: DoS equalling 0 indicates that the stored energy is at the expected level, while DoS close to $\pm$1 means the stored energy is near the allowed range.
\item Since a GES can deviate from its ideal state (DoS=0) to provide services, the DoS can be used to quantify its current flexility, i.e., a DoS value close to 0 implies a high flexility reservation.
\item Finally, as DoS is a generalized index, it can be used to establish a unified dynamic model for various GESs.
\end{enumerate}

\subsection{Derivation of Dynamic Models}
\subsubsection{Electric Energy Storage (EES)}
Ignoring the charge/discharge efficiency, the dynamic model of an EES is given by
\begin{equation}
\label{eq:EES_physical}
E^{EES}_{i,k+1}=E^{EES}_{i,k}+P^{EES}_{i,k}\Delta t,
\end{equation}
where $E^{EES}_{i,k}$ and $P^{EES}_{i,k}$ denote the electric energy and power of EES $i$ at time $k$, respectively ($P^{EES}_{i,k}>0$ when EES is charging); $\Delta t$ is the control cycle.

Variable $S$ is hereafter used to denote DoS. Definition of EES's DoS is given by
\begin{equation}
\label{eq:EES_S}
S^{EES}=-2SOC^{EES}+1.
\end{equation}

Substituting  Eq.\eqref{eq:EES_S} and $SOC^{EES}=E^{EES}/C^{EES}$ into Eq.\eqref{eq:EES_physical}, the following dynamic model can be derived:
\begin{equation}
\label{eq:EES_dynamic}
P^{EES}_{i,k}=-\frac{C^{EES}_i}{2\Delta t}S_{i,k+1}+\frac{C^{EES}_i}{2\Delta t}S_{i,k},
\end{equation}
where $C^{EES}_{i}$ denotes the nominal capacity of EES $i$.

\subsubsection{Electric Vehicle (EV)}
The physical model of an EV is
\begin{equation}
\label{eq:EV_physical}
E^{EV}_{i,k+1}=E^{EV}_{i,k}+\eta^{EV}_{i,k}P^{EV}_{i,k}\Delta t,
\end{equation}
where $E^{EV}_{i,k}$ and $P^{EV}_{i,k}$ denote the electric energy and power of EV $i$ at time $k$, respectively; $\eta^{EV}_{i,k}$ is the charge efficiency.

According to the current charger technology, we focus on the prevailing EV type, which operates at two discrete states: idle and charging with a fixed rate \cite{Binetti2015Scalable,Floch2016PDE}. Let $E_{tar,i}$ denote the target energy of EV $i$ at user-specified departure time $t_{dep,i}$, and $E_{in,i}$ denote the energy at the time EV $i$ is connected into power grid $t_{in,i}$. The average power required to charge EV $i$ to $E_{tar,i}$ can be calculated by
\begin{equation}
\label{eq:EV_Preq}
P_{req,i}=\frac{E_{tar,i}-E_{in,i}}{\eta^{EV}_i(t_{dep,i}-t_{in,i})}.
\end{equation}

If EV $i$ charges at $P_{req,i}$, the expected energy profile is
\begin{equation}
\label{eq:EV_Eexp}
E^{exp}_{i,k+1}=E^{exp}_{i,k}+\eta^{EV}_iP_{req,i}\Delta t.
\end{equation}

An EV can provide its flexibility by deviating from $E^{exp}$. Referring to the hysteretic model in \cite{Duncan2010Achieving}, DoS of EV is defined as
\begin{equation}
\label{eq:EV_S}
S^{EV}_{i,k}=-\frac{E^{EV}_{i,k}-E^{exp}_{i,k}}{C^{EV}_{i}\times r_{i}\%},
\end{equation}
where $C^{EV}_{i}$ denotes the nominal capcity of EV $i$; $r_{i}$ denotes the energy deadband, which limits the error between $E_{tar,i}$ and the actual energy at $t_{dep,i}$ within $\pm C^{EV}_{i}\times r_{i}\%$.

Combined with Eq.\eqref{eq:EV_Preq}-\eqref{eq:EV_S}, Eq.\eqref{eq:EV_physical} can be transformed into
\begin{equation}
\label{eq:EV_dynamic}
P^{EV}_{i,k}=-\frac{C^{EV}_{i}}{\eta^{EV}_{i}\Delta t}S_{i,k+1}+\frac{C^{EV}_{i}}{\eta^{EV}_{i}\Delta t}S_{i,k}+P_{req,i}.
\end{equation}

\subsubsection{Inverter Air-conditioner (IVA)}
Without loss of generality, cooling air-conditioners are studied in this paper. The thermal dynamic process is modelled by a first-order differential equation \cite{Hao2014Aggregate,Song2017Thermal} as
\begin{equation}
\label{eq:IVA_thermal}
\dot{T}_{a,i,k}=-a_i(T_{a,i,k}-T_{o,i,k})-\frac{1}{C_{th,i}}Q^{IVA}_{i,k},
\end{equation}
where $T_{a,i,k}$ and $T_{o,i,k}$ denote the indoor air temperature and outdoor temperature at time $k$, respectively; $Q^{IVA}_{i,k}$ is the heat rate of IVA $i$; the thermal parameter $a_{i}=1/(R_{th,i}C_{th,i})$.

The analytical solution of Eq.\eqref{eq:IVA_thermal} in recursive form is
\begin{equation}
\label{eq:IVA_thermal_recursive}
T_{a,i,k+1}=(T_{a,i,k}-T_{o,i,k}+\frac{1}{a_iC_{th,i}}Q^{IVA}_{i,k})e^{-a_i\Delta t}+T_{o,i,k}-\frac{1}{a_iC_{th,i}}Q^{IVA}_{i,k}.
\end{equation}

The electrical model of an IVA adopts the simplified linear model in \cite{Song2017Thermal}, which is given by
\begin{equation}
\label{eq:IVA_linear}
\left\{
	\begin{array}{lr}
	P^{IVA}_{i}=p_1f_i+p_2\\
	Q^{IVA}_{i}=q_1f_i+q_2
	\end{array}
\right.,
\end{equation}
where $P^{IVA}_i$ denotes the electric power of IVA $i$; $f_i$ is the operation frequency of the compressor; $p_1$, $p_2$, $q_1$ and $q_2$ are coefficients.

DoS of an IVA is defined as
\begin{equation}
\label{eq:IVA_S}
S_{i,k}=\frac{T_{a,i,k}-T_{set,i}}{T_{dev,i}},
\end{equation}
where $T_{set,i}$ denotes the setpoint; $T_{dev,i}$ denotes the allowed temperature deviation.

Combined with Eq.\eqref{eq:IVA_linear} and Eq.\eqref{eq:IVA_S}, Eq.\eqref{eq:IVA_thermal_recursive} can be transformed into
\begin{equation}
\label{eq:IVA_dynamic}
P^{IVA}_{i,k}=\frac{-T_{dev,i}}{\beta_i(1-\alpha_i)}S_{i,k+1}+\frac{\alpha_iT_{dev,i}}{\beta_i(1-\alpha_i)}S_{i,k}+\frac{T_{o,i,k}-T_{set,i}+\gamma_i}{\beta_i},
\end{equation}
where $\alpha_i=e^{-a_i\Delta t}$, $\beta_i=q_1/(a_iC_{th,i}p_1)$, $\gamma_i=(p_1q_2-p_2q_1)/(a_iC_{th,i}p_1)$.

\subsubsection{Fixed-Frequency Air-conditioner (FFA)}
The thermal model of an FFA can also be established by Eq.\eqref{eq:IVA_thermal}. Its electrical model is given by
\begin{equation}
\label{eq:FFA_electric}
Q^{FFA}_{i,k}=COP_i\times P^{FFA}_{i,k},
\end{equation}
where $COP_i$ is the coefficient of performance. FFA is ON/OFF controlled load, so its electric power $P^{FFA}_i$ equals to nominal power $P^{FFA}_{N,i}$ when it is ON and $P^{FFA}_i=0$ when it is OFF.

FFA's definition of DoS is identical to that of IVA, which is given by Eq.\eqref{eq:IVA_S}. Combined with Eq.\eqref{eq:IVA_S} and Eq.\eqref{eq:FFA_electric}, Eq.\eqref{eq:IVA_thermal_recursive} can be transformed to
\begin{equation}
\label{eq:FFA_dynamic}
P^{FFA}_{i,k}=\frac{-T_{dev,i}}{\beta^{'}_i(1-\alpha_i)}S_{i,k+1}+\frac{\alpha_iT_{dev,i}}{\beta^{'}_i(1-\alpha_i)}S_{i,k}+\frac{T_{o,i,k}-T_{set,i}}{\beta^{'}_i},
\end{equation}
where $\beta^{'}_i=R_{th,i}\times COP_i$.

\subsection{Unified Dynamic Model}
The dynamic models of the above GESs can now be represented in a unified form:
\begin{equation}
\label{eq:unified_dynamic}
P_{i,k}=m_{i,1}S_{i,k+1}+m_{i,2}S_{i,k}+m_{i,3},
\end{equation}
where $m_{i,1}$, $m_{i,2}$, $m_{i,3}$ are coefficients of the unified dynamic model.

\begin{figure}[htb]
\centering
\includegraphics[width=15 cm]{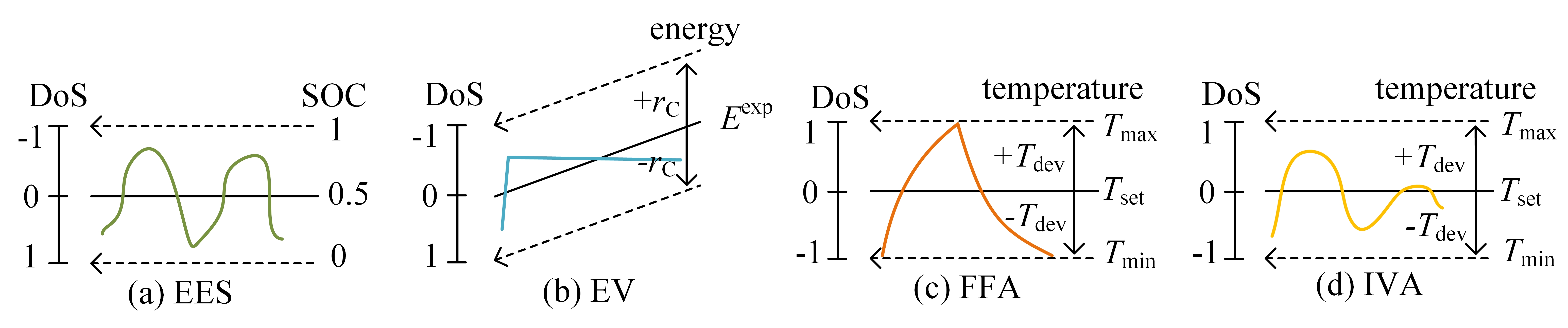}
\caption{ DoS of different GESs.}
\label{fig:Dos_of_GES}
\end{figure}

 Mappings between DoS and the original variables of different GESs are summarized in Fig.\ref{fig:Dos_of_GES}. This paper would study a real-time coordination method of GESs, which could make DoS of individual GESs (or average DoS of a cluster of GESs) approximately equal. Such control characteristic is referred to as DoS-equality control in this paper and will be detailed in the next section.

\section{Real-time Coordination Method of Large-Scale GESs}
\label{sec:coordination_method}
\subsection{ DoS-equality Control Based on Market Equilibrium Mechanism}
Market equilibrium mechanism \cite{Yao2017Transactive} is introduced in this paper to coordinate large-scale GESs. Main stages of the coordination method are:

(1) Bidding: Each GES expresses its urgency and flexibility by constructing a demand curve. The demand curve is denoted by $d_i(\lambda)$ in this paper, which is a non-increasing function.

(2) Aggregating and clearing: LA collects demand curves from all GESs and forms the aggregate demand curve $D(\lambda)=\sum_{i=1}^{N}d_i(\lambda)$, where $N$ is the number of controlled GESs; Assume the aggregate target power is $P_{tar}$, then LA can calculate the clearing price by $\lambda^*=D^{-1}(P_{tar})$.

(3) Disaggregating: LA broadcasts $\lambda^*$ to all GESs. Each GES responds to $\lambda^*$ locally according to its demand curve. The response power of GES $i$ can be obtained by $P_{res,i}=d_i(\lambda^*)$.

Following the above steps, the aggregate target power can be allocated among the GESs, thus realizing accurate power tracking. It should be noted that: (1) The target power $P_{tar}$ depends on applications. In this paper, it will be determined by an optimization problem to be discussed in the next section; (2) The clearing price $\lambda^*$ is only a control signal rather than a price signal. It is dimensionless and its range is set to [-1,1]. Therefore, it is called "virtual price" in this paper.

The proposed market-based coordination method highlights the following advantages:
\begin{enumerate}
\item It improves the autonomy of the GES. Each GES can convert its private information, e.g., user preferences, current adjustable range and security constraints, into demand curve. Since demand curves of all GESs have a unified form, it can shield the differences among various GESs and effectively protect user privacy. Besides, the LA does not have permission to directly control GES, which improves device security.
\item It simplifies the control of the LA. An LA does not need to specify each GES's type and is able to coordinates various GESs via an identical signal, i.e., the virtual price signal $\lambda^*$, which significantly reduces control complexity and requirement of communication bandwidth.
\end{enumerate}

In addition, the proposed method can realize the DoS-equality control to obtain the following advantages:
\begin{enumerate}
\item GESs could have same degree of user satisfaction regardless of the resource type or capacity, which ensures control fairness. In addition, since DoS reflects a GES's state of energy, the DoS-equality control could avoid some GESs going beyond their adjustable range prematurely, thus better utilizing the regulation ability of a GES cluster.
\item The unique DoS of a GES cluster can be a state variable to derive an aggregate dynamic model, making it possible to treat the whole GES cluster as a virtual storage, which will be detailed in section \ref{sec:aggregate_model}.
\end{enumerate}

According to the operation characteristic, the GES can be further classified into two types: GES operating at continuous power (CP-GES) and GES operating at discrete power with discrete states (DP-GES). A CP-GES, e.g., an EES or an IVA, is able to keep its DoS at the desired value by adjusting its operating power, while for a DP-GES, e.g., an EV, an FFA or an electric heater, its DoS generally fluctuates within the allowed range. Besides, the state switching frequency of a DP-GES should generally be controlled to prolong the device's lifetime.

The key of the proposed coordination method is the construction of demand curves for different GESs, which will be introduced in the following subsections.

\subsection{Demand Curve of CP-GES}
\subsubsection{Demand Curve}

Let $S_i$ denote the current DoS of GES $i$. The following construction principle of demand curve is established in this paper:
\begin{equation}
\label{eq:IVA_curve}
\left\{
\begin{aligned}
d_i(\lambda^*)>P_{CONST,i},\lambda^* < S_i\\
d_i(\lambda^*)=P_{CONST,i},\lambda^* = S_i\\
d_i(\lambda^*)<P_{CONST,i},\lambda^* > S_i
\end{aligned}
\right.,
\end{equation}
where $P_{CONST,i}$ denotes the electric power required to maintain the current $S_i$ over a control cycle.

 Eq.\eqref{eq:IVA_curve} is used to realize the DoS-equality control feature. For explanation, the clearing price $\lambda^*$ is assumed to be constant. When $\lambda^*<S_i$, the response power $P_{res,i}=d_i(\lambda^*)$ is higher than $P_{CONST,i}$, leading to the decrease of $S_i$; when $\lambda^*>S_i$, $P_{res,i}$ is lower than $P_{CONST,i}$, leading to the increase of $S_i$; when $\lambda^*=S_i$, $P_{res,i}$ equals $P_{CONST,i}$, keeping $S_i$ unchanged. Therefore, $S_i$ values of all GESs following the principle in Eq.\eqref{eq:IVA_curve} will approach the control signal $\lambda^*$.

\begin{figure}[H]
\centering
\includegraphics[]{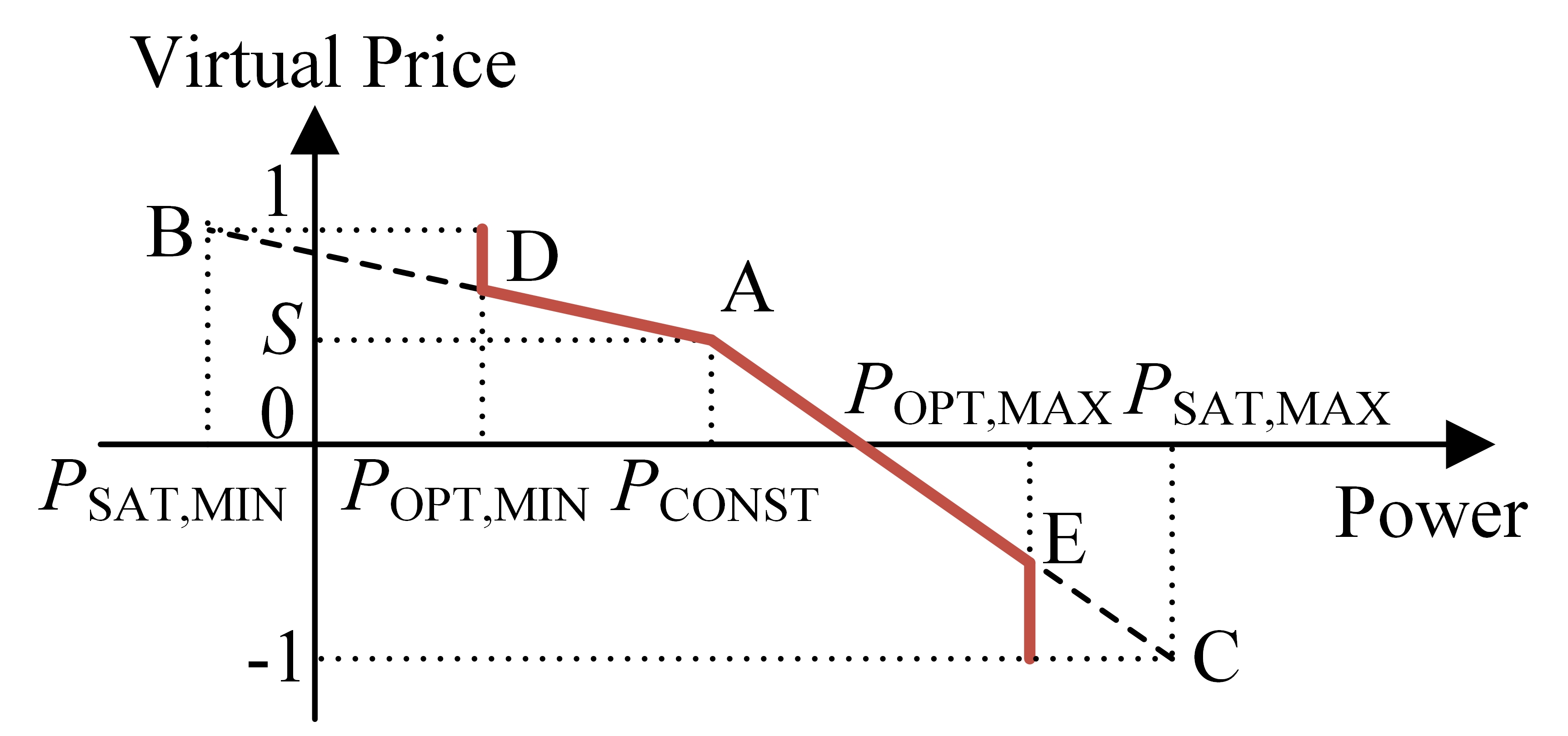}
\caption{ Demand curve of a CP-GES.}
\label{fig:IVA_curve}
\end{figure}

Based on Eq.\eqref{eq:IVA_curve}, the demand curve of CP-GES is constructed as shown in Fig.\ref{fig:IVA_curve}. Note again that the price herein is virtual and is limited between -1 and 1. The demand curve consists of 5 key points. The anchor point A$(P_{CONST},S)$ could satisfy the condition in Eq.\eqref{eq:IVA_curve}. Points B$(P_{SAT,MIN},1)$ and C$(P_{SAT,MAX},-1)$ are used to keep the DoS within limits when responding to any clearing price $\lambda^*$. Thus, $P_{SAT,MIN}$ and $P_{SAT,MAX}$ are minimum and maximum power that would not make DoS go beyond limit over a certain period of time $t_p$ (which is set 5min in this paper), without considering operation constraints. Points D and E which lies on line AB and line AC respectively are further introduced to guarantee the response power would not exceed operational constraints. Thus, $P_{OPT,MIN}$ and $P_{OPT,MAX}$ are minimum and maximum power that a GES could operate at in current control cycle. The calculation method of the above characteristic power will be detailed in the next section.

\subsubsection{Characteristic Power}
\label{sec:CP_characteristic}
The EES and IVA are typical CP-GESs. They can both adopt the demand curve shown in Fig.\ref{fig:IVA_curve}.

An EES's characteristic power can be calculated by

\begin{equation}
\label{eq:EES_characteristic}
\left\{
\begin{aligned}
&P_{CONST}&=& 0\\
&P_{SAT,MIN}&=& \eta^{EES}_{disc}\frac{E^{EES}_{min}-E^{EES}_t}{t_p}\\
&P_{SAT,MAX}&=& \frac{E^{EES}_{max}-E^{EES}_t}{\eta^{EES}_{cha}t_p}\\
&P_{OPT,MIN}&= &-P^{EES}_N\\
&P_{OPT,MAX}&=& P^{EES}_N
\end{aligned}
\right.,
\end{equation}
where $P^{EES}_N$ denotes the nominal charging/discharging power; $\eta^{EES}_{char}$ and $\eta^{EES}_{disc}$ are charge/discharge efficiency, respectively; $E^{EES}_{min}$ and $E^{EES}_{max}$ are the minimum and maximum allowed energy, respectively; $E^{EES}_t$ is the stored energy at time $t$.

IVA's characteristic power can be calculated by
\begin{equation}
\label{eq:IVA_characteristic}
\left\{
\begin{aligned}
&P_{CONST}&=&g^{IVA}_P(T_{a,t_{current}},t_p)\\
&P_{SAT,MIN}&=&g^{IVA}_P(T_{set}+T_{dev},t_p)\\
&P_{SAT,MAX}&=&g^{IVA}_P(T_{set}-T_{dev},t_p)\\
&P_{OPT,MIN}&=&P^{IVA}_{min}\\
&P_{OPT,MAX}&=&P^{IVA}_{max}
\end{aligned}
\right.,
\end{equation}
where $P^{IVA}_{min}$ and $P^{IVA}_{max}$ denotes the minimum and maximum power; the function $g^{IVA}_P(T_{tar},t_p)$ obtains the electric power that makes the indoor air temperature change from current value to the target temperature $T_{tar}$ over a period of time $t_p$. The derivation of $g^{IVA}_P(T_{tar},t_p)$ is detailed in Appendix \ref{sec:appendix_gP}.

\subsection{Demand Curve of DP-GES}
\label{sec:DP_curve}

\subsubsection{Demand Curve}

\begin{figure}[htbp]
\centering
\includegraphics[]{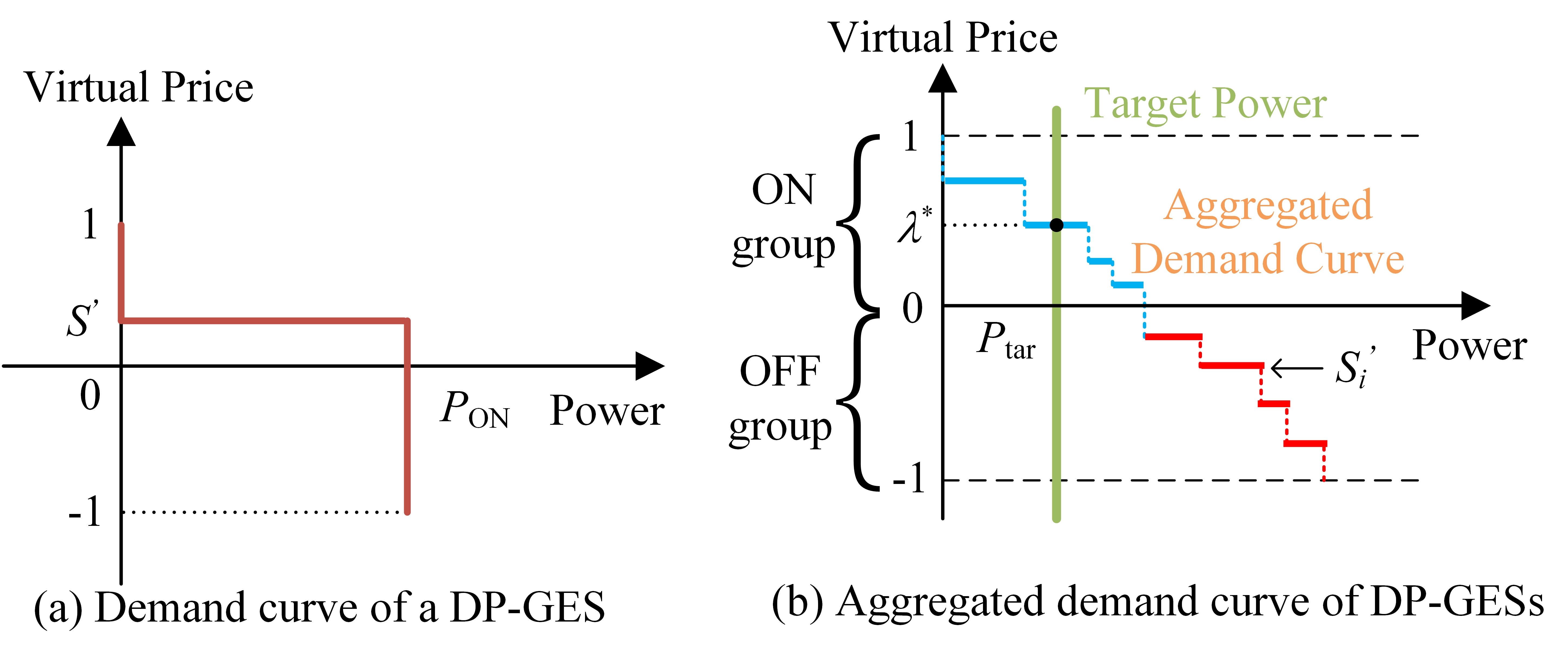}
\caption{Demand curve of DP-GES.}
\label{fig:DPGES_curve}
\end{figure}

A DP-GES typically has two states, i.e., ON and OFF states. The proposed demand curve for a DP-GES is illustrated in Fig.\ref{fig:DPGES_curve}(a), where $P_{ON}$ denotes the operating power when the GES is ON, and $S'$ is transformed from its DoS value by offsetting and then normalizing:
\begin{equation}
\label{eq:Sm}
S_{i}'=\left\{
\begin{aligned}
\frac{S_i+1}{2}&\in[0,1],&state\,ON\\
\frac{S_i-1}{2}&\in[-1,0],&state\,OFF\\
\end{aligned}
\right..
\end{equation}

To explain the principle, the aggregate demand curve of a cluster of DP-GESs is shown in Fig.\ref{fig:DPGES_curve}(b), which is obtained by sorting DP-GESs in descending order of $S'$. As can be seen, the proposed bidding strategy can divide DP-GESs into two groups according to their operation states, i.e., an ON group in the upper half-plane and an OFF group in the lower half-plane, and achieve the following purposes:

(1) For DP-GESs in the same group, a DP-GES's $S'$ reflects its power consumption priority. The higher $S'$ is, the higher probability to maintain or switch to ON state is, and vice versa.

(2) A DP-GES in the ON group always has a higher $S'$ than that in the OFF group, which gives high priority for DP-GESs to maintain their current states, thus avoiding frequent switching.

\begin{figure}[htbp]
\centering
\includegraphics[width=10 cm]{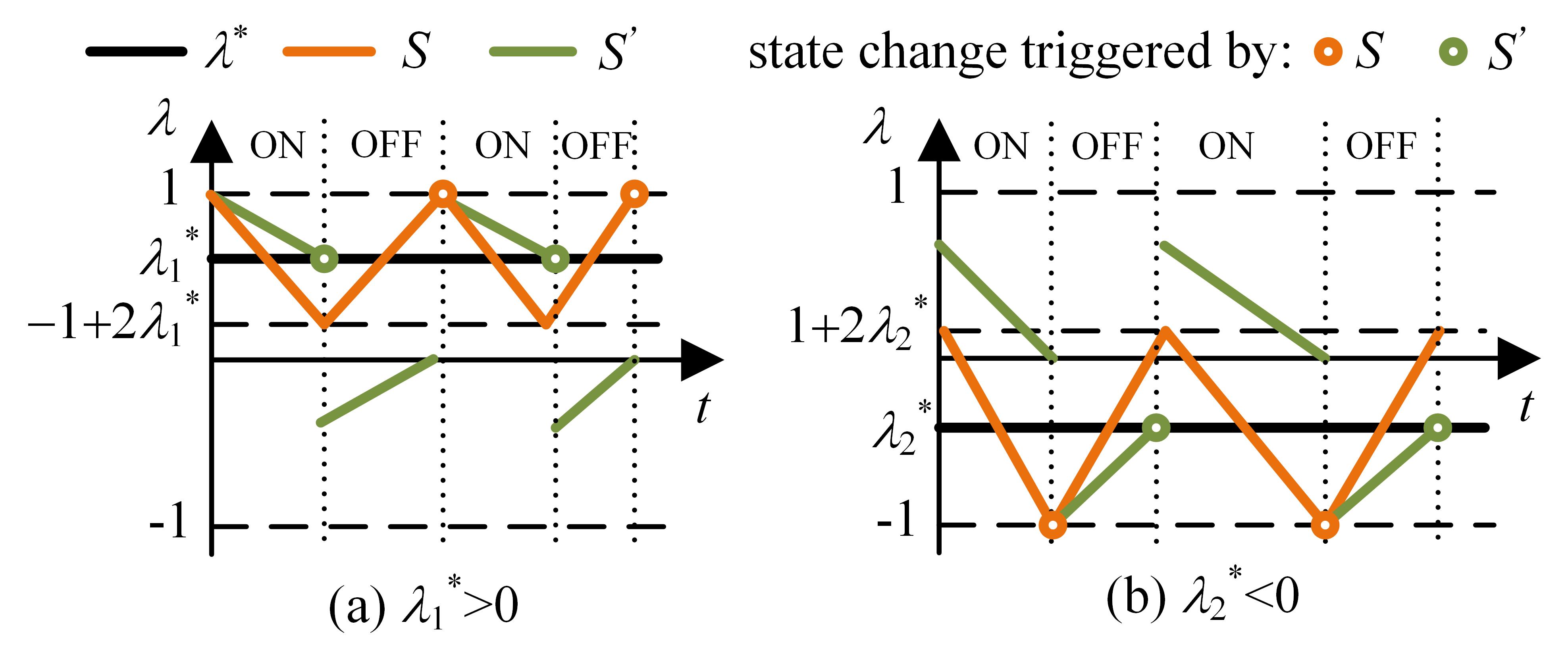}
\caption{$S$ and $S'$ of a DP-GES at different $\lambda^*$.}
\label{fig:DPGES_DoS_Sm}
\end{figure}

To explain the change rule of a single DP-GES's DoS, the clearing price $\lambda^*$ is assumed to be constant over a period of time. Note that the state change of a DP-GES may be triggered either by its $S'$ value, or by $S$ to ensure comfort. The trajectory of $S$ and $S'$ at two different $\lambda^*$, i.e., $\lambda^*_1>0$ and $\lambda^*_2<0$, are illustrated in Fig.\ref{fig:DPGES_DoS_Sm}(a) and Fig.\ref{fig:DPGES_DoS_Sm}(b), and the following laws can be found: when $\lambda^*>0$, $S$ ranges in [-1+2$\lambda^*$,1]; when $\lambda^*<0$, $S$ ranges in [-1,1+2$\lambda^*$], which means that the DoS value fluctuates within a symmetrical range around $\lambda^*$. Therefore, if DoS values of DP-GESs are assumed to be uniformly distributed in such range, the average DoS of the cluster equals $\lambda^*$.

\subsubsection{Characteristic Power}
EV and FFA are typical DP-GESs. They can both adopt the demand curve in Fig.\ref{fig:DPGES_curve}(a).

An EV's characteristic power is
\begin{equation}
\label{eq:EV_characterictic}
P_{ON}=P^{EV}_{N},
\end{equation}
where $P^{EV}_N$ denotes the EV's nominal charging power.

An FFA's characteristic power is
\begin{equation}
\label{eq:FFA_characterictic}
P_{ON}=P^{FFA}_{N}.
\end{equation}

\subsection{Locked State}
In certain control cycles, a GES may get into the locked state, which means it should maintain its current operating state and power. There are two typical situations: (1) To reduce mechanical wear and protect the device \cite{Liu2016Model,Tang2017A}, DP-GESs, such as FFA and EV, should satisfy lockout time constraints before they switch state; (2) Restricted by the device capability, response cycle of some GESs may be longer than the real-time control cycle (which is 10s in this paper).

The lockout mechanism can be easily realized in this paper, as a GES can simply submit the following demand curve during the lockout time:
\begin{equation}
\label{eq:locked_curve}
d_i(\lambda)=P_{i,t},\forall \lambda,
\end{equation}
which means the response power of GES $i$ maintains its current operating power $P_{i,t}$ for any $\lambda$.

\section{Aggregate Dynamic Model of a GES Cluster}
\label{sec:aggregate_model}

One significant advantage of our method is that the aggregate dynamic model of a large-scale GES cluster can be easily derived thanks to the DoS-equality control feature, as well as the unified dynamic model for different GESs defined in Eq.\eqref{eq:unified_dynamic}.

For CP-GESs, add their dynamic models together:
\begin{equation}
\label{eq:CPGES_dynamic_add}
\sum_{i\in\Omega_c}P_{i,k}=\sum_{i\in\Omega_c}m_{i,1}S_{i,k+1}+\sum_{i\in\Omega_c}m_{i,2}S_{i,k}+\sum_{i\in\Omega_c}m_{i,3},
\end{equation}
where $\Omega_c$ represents the set of CP-GESs.

Under the DoS-equality control, DoS of all CP-GESs becomes equal. Let $S_{agg-c}$ denote DoS of the CP-GES cluster, and $P_{agg-c}$ denote the aggregate power. The aggregate dynamic model of CP-GESs can then be derived as
\begin{equation}
\label{eq:CPGES_Ag_dynamic}
P_{agg-c,k}=S_{agg-c,k+1}\sum_{i\in\Omega_c}m_{i,1}+S_{agg-c,k}\sum_{i\in\Omega_c}m_{i,2}+\sum_{i\in\Omega_c}m_{i,3}.
\end{equation}

For DP-GESs, to facilitate analysis, we assume coefficients of their dynamic models to be equal first. Add their dynamic models together, and we obtain
\begin{equation}
\label{eq:DPGES_dynamic_add}
\sum_{i\in\Omega_d}P_{i,k}=m_{i,1}\sum_{i\in\Omega_d}S_{i,k+1}+m_{i,2}\sum_{i\in\Omega_d}S_{i,k}+\sum_{i\in\Omega_d}m_{i,3},
\end{equation}
where $\Omega_d$ represents the set of DP-GESs.

The instantaneous DoS value of each DP-GES is a random variable. Denote the average DoS of DP-GESs as $S_{agg-d}$, and denote the aggregate power as $P_{agg-d}$. Then Eq.\eqref{eq:DPGES_dynamic_add} can be written as:
\begin{equation}
\label{eq:DPGES_dynamic_add2}
P_{agg-d,k}=|\Omega_d|m_{i,1}S_{agg-d,k+1}+|\Omega_d|m_{i,2}S_{agg-d,k}+|\Omega_d|m_{i,3},
\end{equation}
where $|\Omega_d|$ denotes the number of DP-GESs.

Under the DoS-equality control, $S_{agg-c}$ of CP-GESs and $S_{agg-d}$ of DP-GESs tend to be equal (both equal $\lambda^*$), thus can both be denoted by $S_{agg}$. Therefore, Eq.\eqref{eq:CPGES_Ag_dynamic} and Eq.\eqref{eq:DPGES_dynamic_add2} can be added together:
\begin{equation}
\label{eq:GES_Ag_dynamic}
P_{agg,k}=S_{agg,k+1}\sum^N_{i=1}m_{i,1}+S_{agg,k}\sum^N_{i=1}m_{i,2}+\sum^N_{i=1}m_{i,3},
\end{equation}
where $N=|\Omega_c|+|\Omega_d|$ denotes the total number of GESs; $P_{agg,k}$ is the aggregate power.

Eq.\eqref{eq:GES_Ag_dynamic} can be represented in a compact form as
\begin{equation}
\label{eq:GES_Ag_dynamic2}
P_{agg,k}=M_{1,k}S_{agg,k+1}+M_{2,k}S_{agg,k}+M_{3,k}.
\end{equation}

The above derivation is based on the assumption that all DP-GESs have equal model coefficients. For those with different coefficients, we can first divide them into different groups according to their coefficients, then aggregate each group using Eq.\eqref{eq:DPGES_dynamic_add2}, and finally form the aggregate dynamic model of all the GESs (including CP-GESs and DP-GESs) by Eq.\eqref{eq:GES_Ag_dynamic}.

The aggregate dynamic model of heterogeneous GESs proposed above has the advantage that the LA could obtain the aggregate model easily by adding model coefficients of all GESs with no need to identify GES's type, thus having a low computational cost. Furthermore, from the control aspect, the low-dimensional aggregate model greatly reduces the complexity of the optimization problem, which will be discussed in the next section.

\section{Application}
\label{sec:application}

\subsection{Optimal Multi-Market Flexibility Allocation}
\begin{figure}[H]
\centering
\includegraphics[]{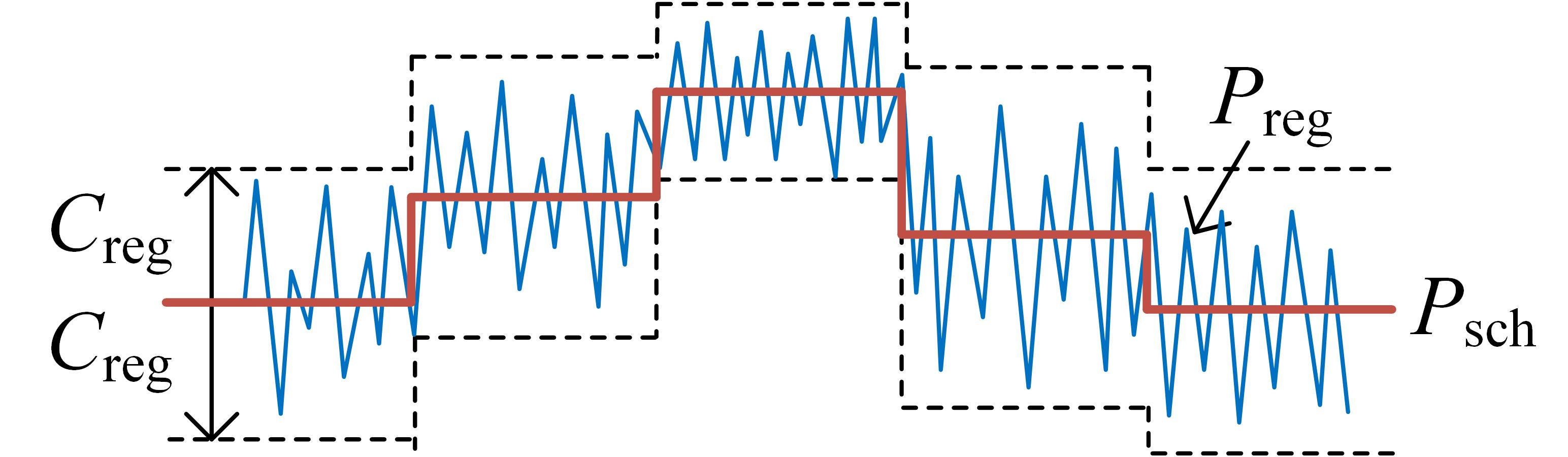}
\caption{Participation in energy and regulation market.}
\label{fig:energy_regulation_market}
\end{figure}

An LA can aggregate flexibility of large-scale GESs to provide multiple services to power grid. For example, it can schedule an optimal power consumption profile according to the electricity price of energy market \cite{Song2017Thermal,Hao2017Optimal,Zhang2016A}, as the profile $P_{sch}$ in Fig.\ref{fig:energy_regulation_market}. LA can provide other ancillary services at the same time to obtain higher benefits \cite{Hao2017Optimal,Zhang2016A}, e.g., responding to the regulation signal $P_{reg}$ in Fig.\ref{fig:energy_regulation_market}.

The regulation signal can be decomposed into the low frequency part denoted by regA, and the high frequency part denoted by regD \cite{Xu2016A}. Literature \cite{Zechun2014Research} analyses the regulation signal of a certain power grid, and finds that the high frequency part could account for up to 30\%. In this paper, the LA responds to the regD signal considering the following facts: First, the regD signal has zero-mean over a period of time \cite{Xu2016A}, which could significantly reduce requirements for the capacity of GESs; Second, the extra energy introduced by the regD signal is close to 0, thus having little impact on electricity bills; Third, if regulation payments are determined by the performance-based policy used in PJM regulation market \cite{Aho2015Controlling}, the LA could obtain high benefits.

Considering both the energy and regulation markets, the target power of an LA is given by
\begin{equation}
\label{eq:Ptar}
P_{tar}=P_{sch}+P_{reg}=P_{sch}+regD\times C_{reg},
\end{equation}
where $P_{sch}$ is the hourly scheduled power, which determines the bill paid to the energy market; regD is the regulation signal normalized to [0,1]; $C_{reg}$ denotes the contracted regulation capacity.

Let the optimization cycle be 1 hour, then the LA solves the following convex optimization problem in the $n$th cycle to allocate its flexibility to the two markets:
\begin{equation}
\label{eq:optimization_problem}
\begin{aligned}
&\min_{P_{sch},C_{reg},S_{agg}}\sum^{24}_{k=n}\overbrace{\mu_{ele,k}P_{sch,k}}^{electricity\;bill}- \overbrace{(\underbrace{\hat{\omega}_{score}\mu_{cap,k}C_{reg,k}}_{capacity\;payments}+\underbrace{\hat{\omega}_{score}\mu_{mile,k}\hat{\omega}_{mile}C_{reg,k}}_{mileage\;payments})}^{regulation\;payments}+f_s(S_{agg,k})\\
s.t.
&P_{sch,k}=M_{1,k}S_{agg,k+1}+M_{2,k}S_{agg,k}+M_{3,k},\forall k\\
&-1\le S_{agg,k}\le 1,\forall k\\
&P_{sch,k}+C_{reg,k}\le P^{max}_{agg,k},\forall k\\
&P_{sch,k}-C_{reg,k}\ge P^{min}_{agg,k},\forall k\\
&C_{reg,k}\ge 0,\forall k,
\end{aligned}
\end{equation}
where $\mu_{ele,k}$, $\mu_{cap,k}$, $\mu_{mile,k}$ denote respectively the electricity price, regulation capacity price and regulation mileage price in the $k$th cycle; $\hat{\omega}_{score}$ denotes the statistical value of the regulation performance score defined by PJM \cite{Aho2015Controlling}, and $\hat{\omega}_{mile}$ denotes the statistical value of the regulation mileage  \cite{Aho2015Controlling}; $P^{max}_{agg,k}$ and $P^{min}_{agg,k}$ are the maximum and minimum power of the GES cluster, which are calculated in every optimization cycle as $P^{max}_{agg}=\sum_{i=1}^NP_{max,i}$ and $P^{min}_{agg}=\sum_{i=1}^NP_{min,i}$, where
\begin{equation}
\label{eq:PminPmax}
\begin{aligned}
P_{max,i}&=\left\{
\begin{aligned}
&P_{OPT,MAX,i}&,i\in\Omega_c\\
&P_{ON,i}&,i\in\Omega_d
\end{aligned}
\right.,\\
P_{min,i}&=\left\{
\begin{aligned}
&P_{OPT,MIN,i}&,i\in\Omega_c\\
&0&,i\in\Omega_d
\end{aligned}
\right..
\end{aligned}
\end{equation}
Note that EESs are able to discharge, thus the LA may sell electricity to power grid at this time. The purchase price and sale price are assumed to be equal in this paper.

The first constraint in Eq.\eqref{eq:optimization_problem} is the aggregate dynamic model of the GES cluster, the second constraint ensures the user satisfaction, and the last three ones constrain the regulation capacity. Thanks to the established aggregate dynamic model, the scale of the GES cluster does not affect the computational complexity of the optimization problem.

To improve user satisfaction, Eq.\eqref{eq:optimization_problem} includes a penalty term $f_s(S_{agg,k})$, which is defined as
\begin{equation}
\label{eq:fs}
f_s(S_{agg,k})=\omega_{scale}\mu^{avg}_{ele}S^2_{agg,k}(P^{max}_{agg,k}-P^{min}_{agg,k}),
\end{equation}
where $\omega_{scale}$ is a proportionality coefficient, which is assigned 0.1 here; $\mu^{avg}_{ele}$ is the daily average electricity price.

\subsection{Three-Layer Control Structure}
\begin{figure}[H]
\centering
\includegraphics[]{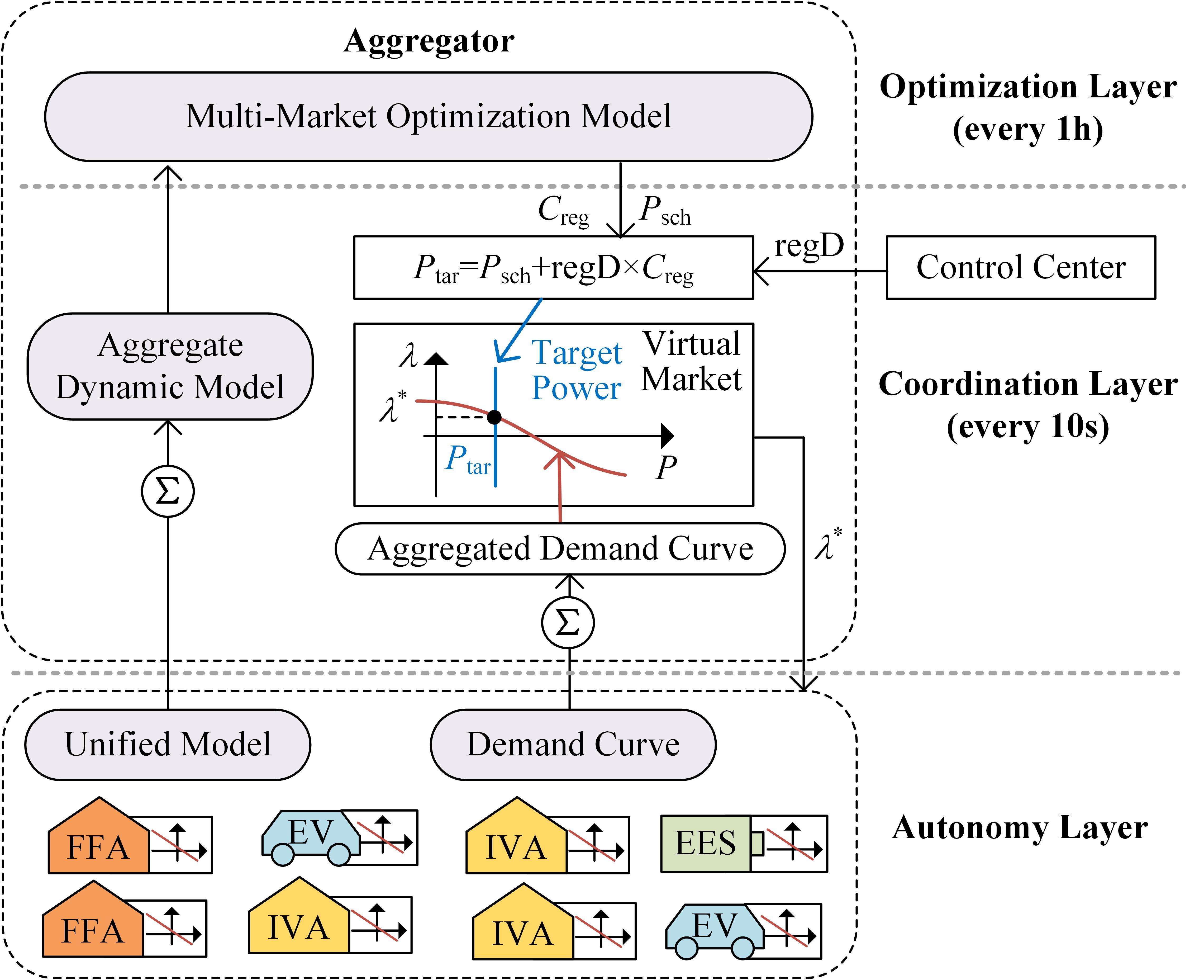}
\caption{Three-layer control structure.}
\label{fig:three_layer}
\end{figure}

To sum up, this paper develops four models, i.e., a unified dynamic model and a unified demand model for individual GES, as well as an aggregate dynamic model and an optimization model for an LA to approximate dynamics of a GES cluster and participate in multi-markets. These four models are organized in a a three-layer control structure, as illustrated in Fig.\ref{fig:three_layer}.

(1) Rolling Optimization Layer

At the beginning of current optimization cycle $n$, the LA gathers model coefficients $m_{i,1}\sim m_{i,3}$ from GESs and update the aggregate dynamic model according to Eq.\eqref{eq:GES_Ag_dynamic}. Meanwhile, current DoS values of all GESs are collected to update the initial recursive value of $S_{agg,n}$ by
\begin{equation}
\label{eq:update_Sagg}
\hat{S}_{agg,n}=\frac{1}{N}\sum^{N}_{i=1}S_{i,n}
\end{equation}

The optimization problem in Eq.\eqref{eq:optimization_problem} can then be solved, and the optimal scheduled power sequence $[P_{sch,n},...,P_{sch,24}]$ as well as the regulation capacity sequence $[C_{reg,n},...,C_{reg,24}]$ are obtained. Implement only the first elements of these optimal sequences, i.e., $P_{sch,n}$ and $C_{reg,n}$ in the current optimization cycle, and repeat the above steps each hour. It's well known that this idea of rolling optimization comes from the model predictive control (MPC) \cite{Liu2016Model}, which is adopted here to update the aggregate dynamic model iteratively, and to consider constraints in future time slots explicitly.

(2) Real-time Coordination Layer

In each control cycle (10s in this paper), the LA receives the regulation signal regD from the control center, and then calculates the real-time target power $P_{tar}$ according to Eq.\eqref{eq:Ptar}. The virtual market is cleared according to section \ref{sec:coordination_method}, and the clearing price $\lambda^*$ is broadcast to each GES.

(3) GES Autonomy Layer

In each optimization cycle (1h), each GES updates its DoS, model coefficients in Eq.\eqref{eq:unified_dynamic} and power constraints in Eq.\eqref{eq:PminPmax}, and then reports these information to the LA. In each control cycle (10s), each GES reports its flexibility and responds to the clearing price both through the demand curve.

It is worth mentioning that, since the LA interacts with different GESs through a unified set of information, i.e., DoS, model coefficients, power constraints, demand curve and virtual price, the method in this paper supports a flexible tree-like structure. For example, a local concentrator can be deployed in an community to pre-aggregate the information. Therefore, the method has high scalability and is suitable for wide-area coordination of large-scale GESs.

\section{Simulation Studies}
\label{sec:simulation}

\subsection{Simulation Settings}
\begin{figure}[H]
\centering
\includegraphics[width=8 cm]{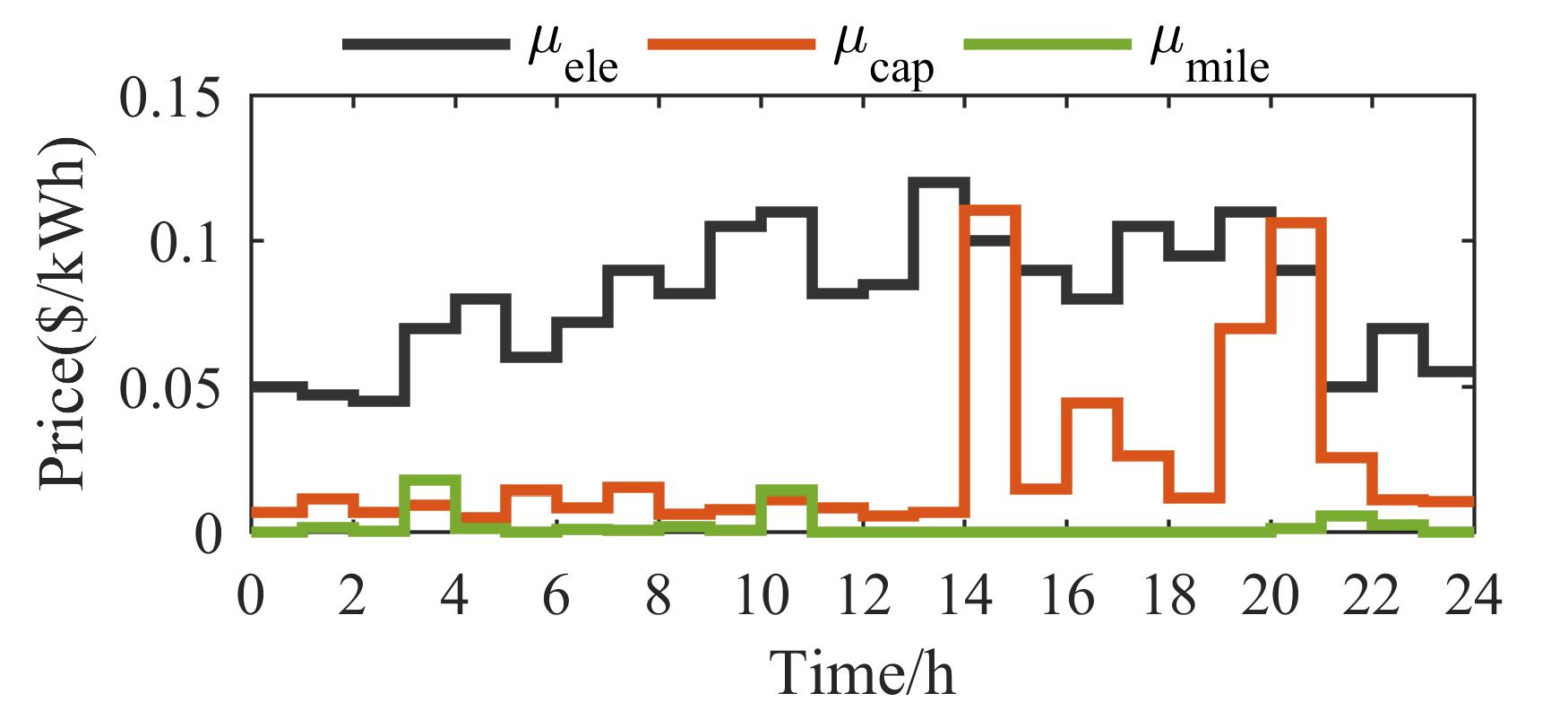}
\caption{Price.}
\label{fig:price}
\end{figure}

The simulation cases are based on a residential community system. The simulation lasts for 24h. Electricity price $\mu_{ele}$, regulation capacity price $\mu_{cap}$ and mileage price $\mu_{mile}$ adopt the data in literature \cite{Song2017Thermal} and \cite{Yao2018A}, and their profile are illustrated in Fig.\ref{fig:price}. The regD signal uses the PJM data on Jul 13th, 2016. According to the statistical analysis on the regD signal in 2016 \cite{PJM_regD_data}, we set $\hat{\omega}_{mile}=2.7$. According to the simulation results based on historical data, we conservatively assign $\hat{\omega}_{score}=0.92$.

Four types of GESs are considered in this paper, i.e., EES, EV, FFA and IVA, whose parameters are shown in Table.\ref{tab:parameter}, where U(a,b) indicates a uniform distribution between [a,b] and $t_{res}$ denotes the response cycle of GES. The profile of the outdoor temperature in simulation cases is shown in Fig.\ref{fig:To}.

\begin{table}[H]
  \centering
  \caption{Parameters of GESs.}
    \begin{tabular}{cccc|cccc}
    \toprule
    \multicolumn{2}{c}{\textbf{Type}} & \textbf{Parameter} & \textbf{Value} & \multicolumn{2}{c}{\textbf{Type}} & \textbf{Parameter} & \textbf{Value} \\

    \midrule
    \multicolumn{2}{c}{\multirow{5}[10]{*}{EES}} & Number & 10    & \multicolumn{1}{c}{\multirow{14}[28]{*}{TCL}} & \multirow{2}[4]{*}{Thermal Parameter} & $R_{th}$($^{\circ}C$/kW) & U(1,1.5) \\

\cmidrule{3-4}\cmidrule{7-8}    \multicolumn{2}{c}{} & $C^{EES}$(kWh) & U(40,50) &       &       & $C_{th}$(kWh/$^{\circ}C$) & U(0.8,1.2) \\

\cmidrule{3-4}\cmidrule{6-8}    \multicolumn{2}{c}{} & $P^{EES}_N$(kW) & U(40,50) &       & \multirow{2}[4]{*}{Preference} & $T_{set}$($^{\circ}C$) & U(23,28) \\

\cmidrule{3-4}\cmidrule{7-8}    \multicolumn{2}{c}{} & $\eta^{EES}_{char}$/$\eta^{EES}_{disc}$ & 0.9/0.9 &       &       & $T_{dev}$($^{\circ}C$) & U(2,3) \\

\cmidrule{3-4}\cmidrule{6-8}    \multicolumn{2}{c}{} & $t_{res}$(s) & 10    &       & \multirow{4}[8]{*}{FFA} & Number & 100 \\

\cmidrule{1-4}\cmidrule{7-8}    \multicolumn{2}{c}{\multirow{9}[18]{*}{EV}} & Number & 20    &       &       & $P^{FFA}_N$(kW) & U(4.5,5.5) \\

\cmidrule{3-4}\cmidrule{7-8}    \multicolumn{2}{c}{} & $C^{EV}$(kWh) & U(20,30) &       &       & COP   & U(3,4) \\

\cmidrule{3-4}\cmidrule{7-8}    \multicolumn{2}{c}{} & $P^{EV}_N$(kW) & U(6,8) &       &       & $t_{lock}$(min) & 5 \\

\cmidrule{3-4}\cmidrule{6-8}    \multicolumn{2}{c}{} & $\eta^{EV}$ & 0.9   &       & \multirow{6}[12]{*}{IVA} & Number & 100 \\

\cmidrule{3-4}\cmidrule{7-8}    \multicolumn{2}{c}{} & $t_{in}$(h) & U(18,22) &       &       & $P^{IVA}_{max}$(kW) & U(5,6) \\

\cmidrule{3-4}\cmidrule{7-8}    \multicolumn{2}{c}{} & $t_{dep}$(h) & U(6,9) &       &       & $P^{IVA}_{min}$(kW) & U(0.4,0.5) \\

\cmidrule{3-4}\cmidrule{7-8}    \multicolumn{2}{c}{} & $r$\%   & 2.50\% &       &       & $p_1$/$q_1$(kW/Hz) & 0.03/0.06 \\

\cmidrule{3-4}\cmidrule{7-8}    \multicolumn{2}{c}{} & $SOC_{tar}$ & U(0.75,0.85) &       &       & $p_2$/$q_2$(kW) & -0.4 / -0.3 \\

\cmidrule{3-4}\cmidrule{7-8}    \multicolumn{2}{c}{} & $t_{lock}$(min) & 5     &       &       & $t_{res}$(s) & 60 \\
    \bottomrule
    \end{tabular}
  \label{tab:parameter}
\end{table}

To evaluate the control effect, the LA has to estimate the baseline power of the GES cluster when all GESs are uncontrolled, which will be denoted by $P_{base}$ hereafter. Many papers have studied the estimation method of the baseline load, e.g., a statistical based method proposed in \cite{Chen2017Short}. In this paper, since the aggregate model of the GES cluster is available to the LA, it can estimate $P_{base}$ in each hour by solving the following optimization problem:
\begin{equation}
\label{eq:optimal_Pbase}
\begin{aligned}
&\min_{P_{base},S_{agg}}\sum^{24}_{k=n}S_{agg,k}^2\\
s.t.
& P_{base,k}=M_{1,k}S_{agg,k+1}+M_{2,k}S_{agg,k}+M_{3,k},\forall k\\
& -1\le S_{agg,k}\le 1,\forall k\\
& P_{agg,min,k}\le P_{base,k}\le P_{agg,max,k},\forall k.\\
\end{aligned}
\end{equation}

Solution of Eq.\eqref{eq:optimal_Pbase} is referred to as the baseline case in this paper.

\subsection{Case 1: only participate in the energy market}
To evaluate the effectiveness of the aggregate dynamic model and the DoS-equality control, an optimization problem simplified from Eq. \eqref{eq:optimization_problem} is solved which only considers the energy market:
\begin{equation}
\label{eq:optimal_EM}
\begin{aligned}
&\min_{P_{sch},S_{agg}}\sum^{24}_{k=n}[\mu_{ele,k}P_{sch,k}+f_s(S_{agg,k})]\\
s.t.
& P_{sch,k}=M_{1,k}S_{agg,k+1}+M_{2,k}S_{agg,k}+M_{3,k},\forall k\\
& -1\le S_{agg,k}\le 1,\forall k\\
& P_{agg,min,k}\le P_{sch,k}\le P_{agg,max,k},\forall k.\\
\end{aligned}
\end{equation}

\begin{figure}[H]
\centering
\includegraphics[width=8 cm]{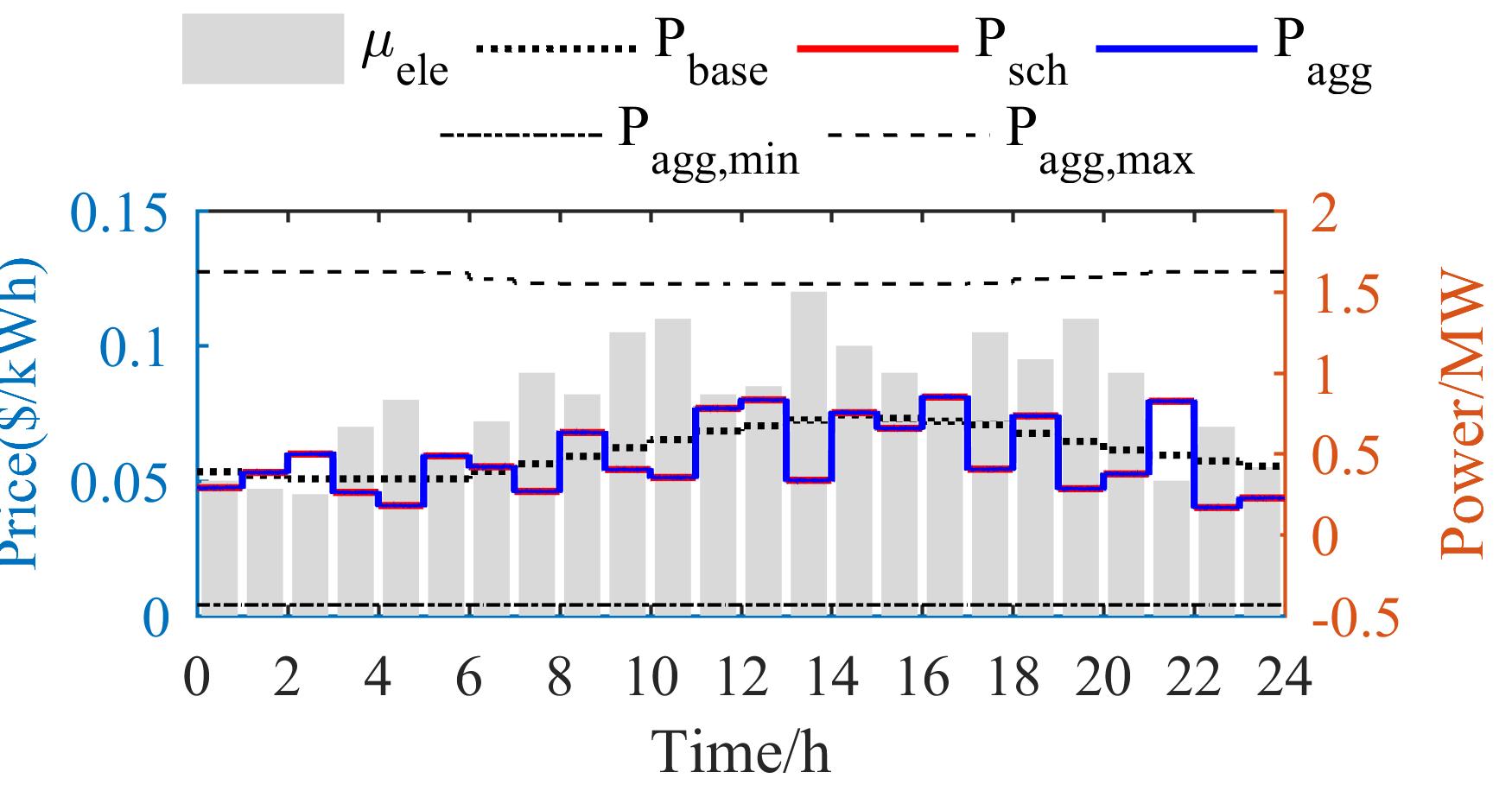}
\caption{Aggregate power and electricity price in case 1.}
\label{fig:EM_P_price}
\end{figure}

LA coordinates GESs to make the aggregate power $P_{agg}$ track the scheduled power $P_{sch}$. As illustrated in Fig.\ref{fig:EM_P_price}, the proposed method has high tracking accuracy. In addition, the aggregate power $P_{agg}$ can vary up and down around the baseline power $P_{base}$ as the electricity price changes to reduce energy cost. Therefore, the GES cluster can be scheduled as virtual energy storage, since it can be charged by making $P_{agg}$ higher than $P_{base}$ and discharged by making $P_{agg}$ lower than $P_{base}$.

\begin{figure}[H]
\centering
\includegraphics[width=8 cm]{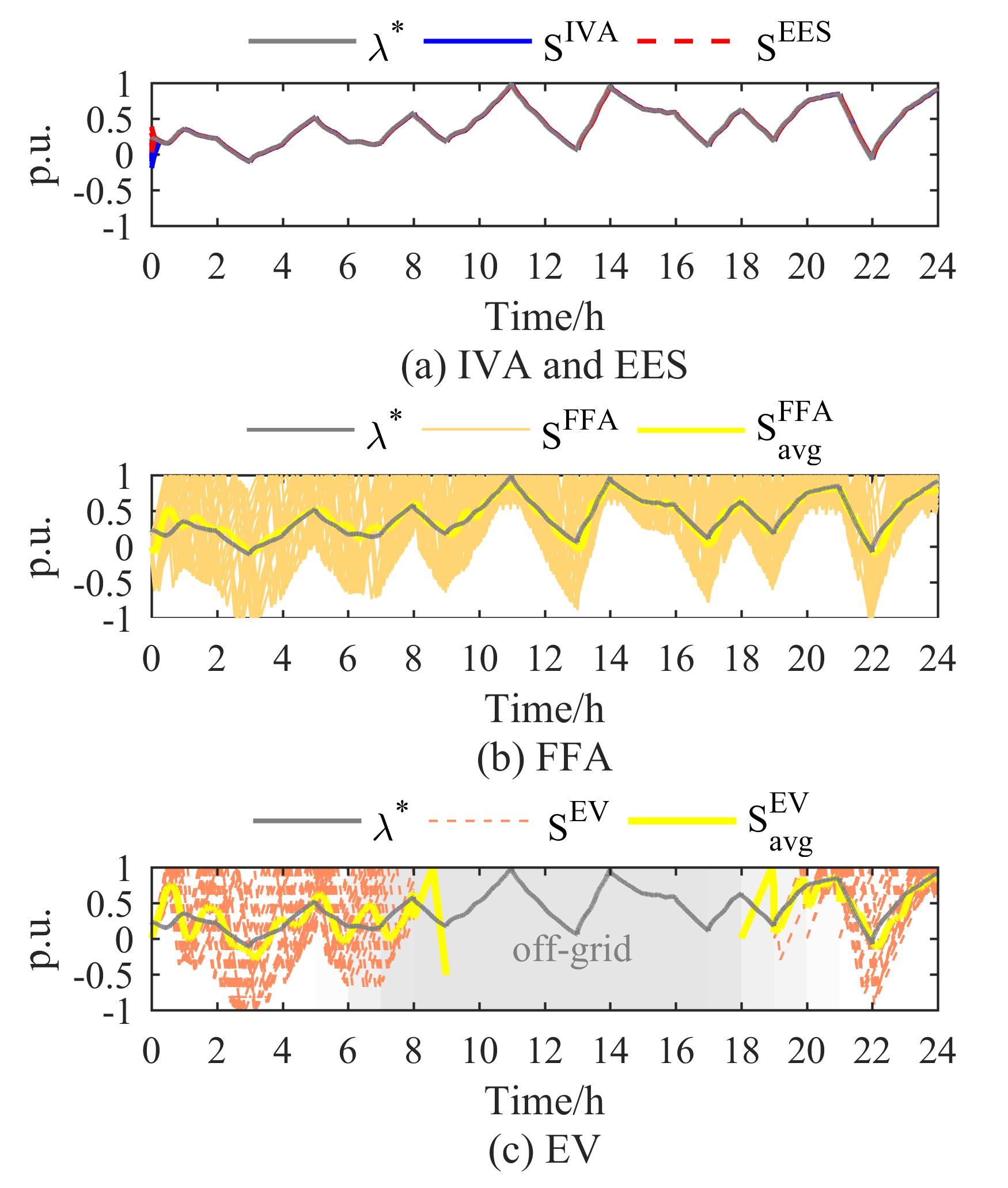}
\caption{Performance of the DoS-equality control in case 1.}
\label{fig:EM_S3}
\end{figure}

Fig.\ref{fig:EM_S3} demonstrates the performance of the DoS-equality control. As shown in Fig.\ref{fig:EM_S3}(a), DoS of all CP-GESs, i.e., IVAs and EESs, can track $\lambda^*$ quite well as they are able to adjust their power continuously. For DP-GESs (including FFAs and EVs), note again that the average DoS of a cluster rather than an individual's DoS can follow $\lambda^*$. It can be seen in Fig.\ref{fig:EM_S3}(b) that $S^{FFA}_{avg}$ of FFAs tracks $\lambda^*$ well. However, for the EVs, $S^{EV}_{avg}$ slightly fluctuates around $\lambda^*$. This is because the number of EVs is small, making the statistical characteristics inconspicuous and the distribution of DoS not well aligned with the analysis in section \ref{sec:DP_curve}. Therefore, a larger-scale DP-GES cluster yields better DoS-equality control effect.

\begin{figure}[H]
\centering
\includegraphics[width=8 cm]{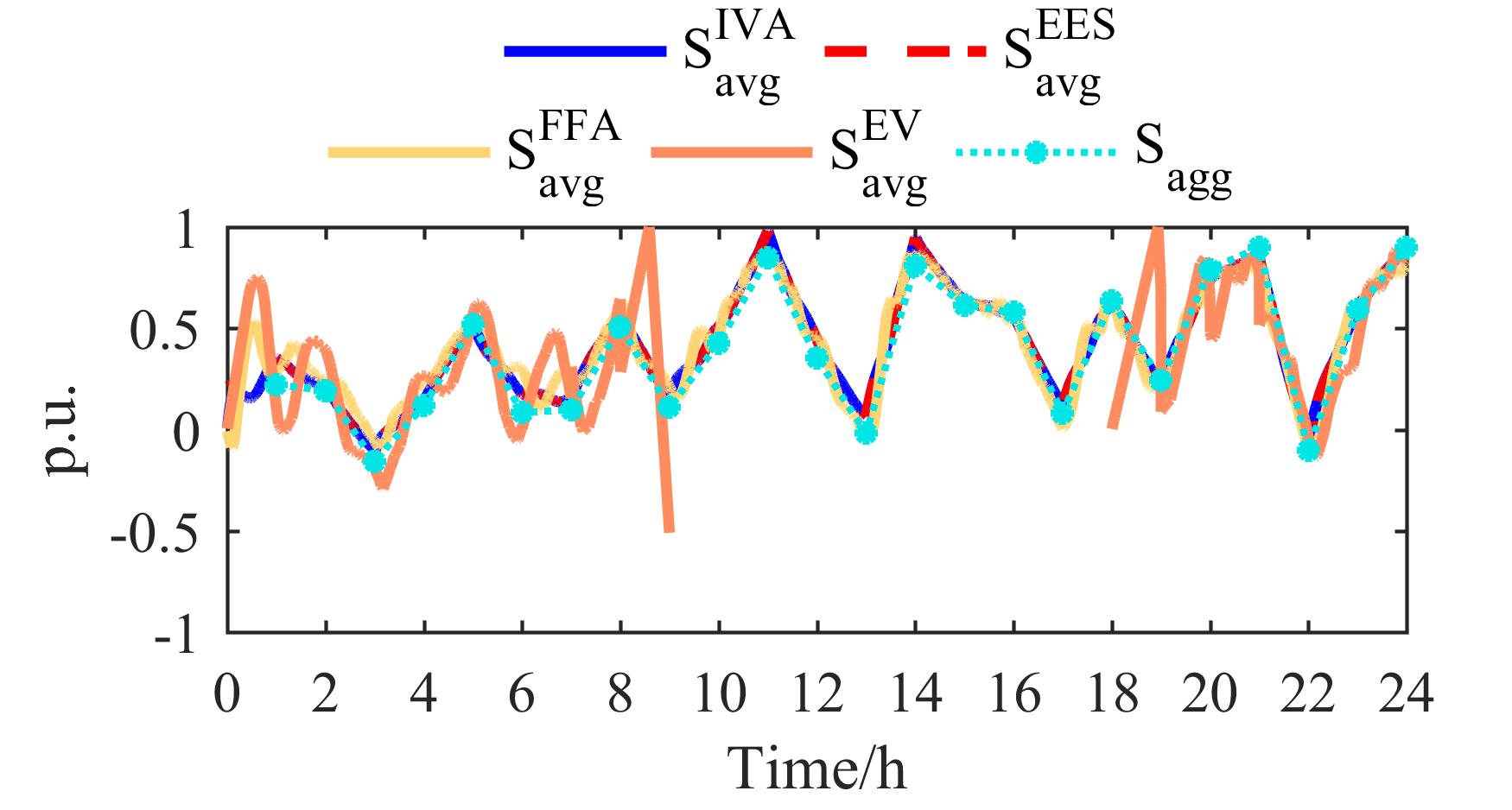}
\caption{Average DoS and aggregate DoS in case 1.}
\label{fig:EM_Savg}
\end{figure}

Fig.\ref{fig:EM_Savg} shows how well the aggregate dynamic model defined in Eq.\eqref{eq:GES_Ag_dynamic2} fits the cluster with heterogeneous GESs. As can be seen, $S_{avg}$ of all GESs except EVs are very close to the aggregate state variable $S_{agg}$ at the end of every optimization cycle. Since the number of EVs is small, especially during 6:00-9:00 and 18:00-20:00 when some EVs are off-grid, $S^{EV}_{avg}$ fluctuates around $S_{agg}$ with relatively large errors.

In addition to the small number of DP-GESs, some other factors may also lead to errors in the aggregate dynamic model, e.g., the charge/discharge efficiency is not considered in EES's dynamic model. In order to prevent the error from being accumulated, this paper adopts the rolling optimization method to mitigate impacts of these factors continually. As can be observed from Fig.\ref{fig:EM_Savg}, when combined with the rolling optimization, the simple aggregate model  in \eqref{eq:GES_Ag_dynamic2} could be a useful tool for an LA to capture the aggregate dynamic feature of large-scale GESs.

\subsection{Case 2: participate in both energy and regulation markets}
\begin{figure}[H]
\centering
\includegraphics[width=8 cm]{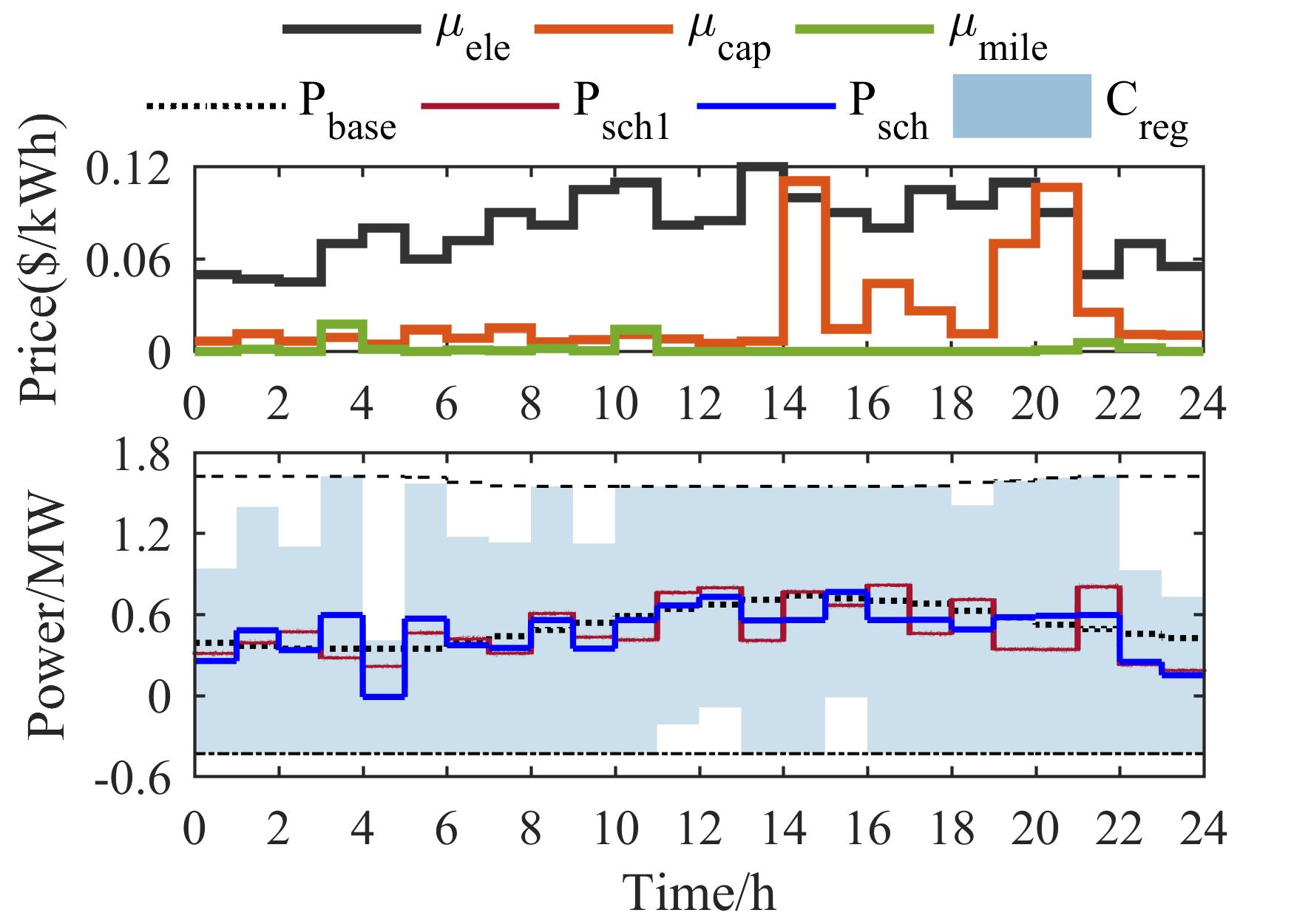}
\caption{Scheduled power and regulation capacity in case 2.}
\label{fig:ER_Psch_C}
\end{figure}

In case 2, the GES cluster participates in both the energy and regulation markets. By solving the optimization problem in Eq.\eqref{eq:optimization_problem}, the scheduled power $P_{sch}$ and regulation capacity $C_{reg}$ can be obtained, as illustrated in Fig.\ref{fig:ER_Psch_C}. For comparison purposes, the scheduled power in case 1 is also plotted in the figure, which is denoted by $P_{sch1}$. The scheduled power profile in case 2 has significant difference from that in case 1 because the LA should allocate the cluster's flexibility to two markets according to both electricity price ($\mu_{ele}$) and regulation price ($\mu_{cap}$ and $\mu_{mile}$). Since a symmetric regulation signal is used in this paper, $C_{reg}$ can reach its maximum value when $P_{sch}\approx (P_{agg,min}+P_{agg,max})/2$. It can be seen that when $\mu_{cap}$ is relatively high, e.g., during 14:00-15:00, 19:00-21:00, the LA tends to maximize $C_{reg}$ to gain higher payments from regulation market.

\begin{figure}[H]
\centering
\includegraphics[width=8 cm]{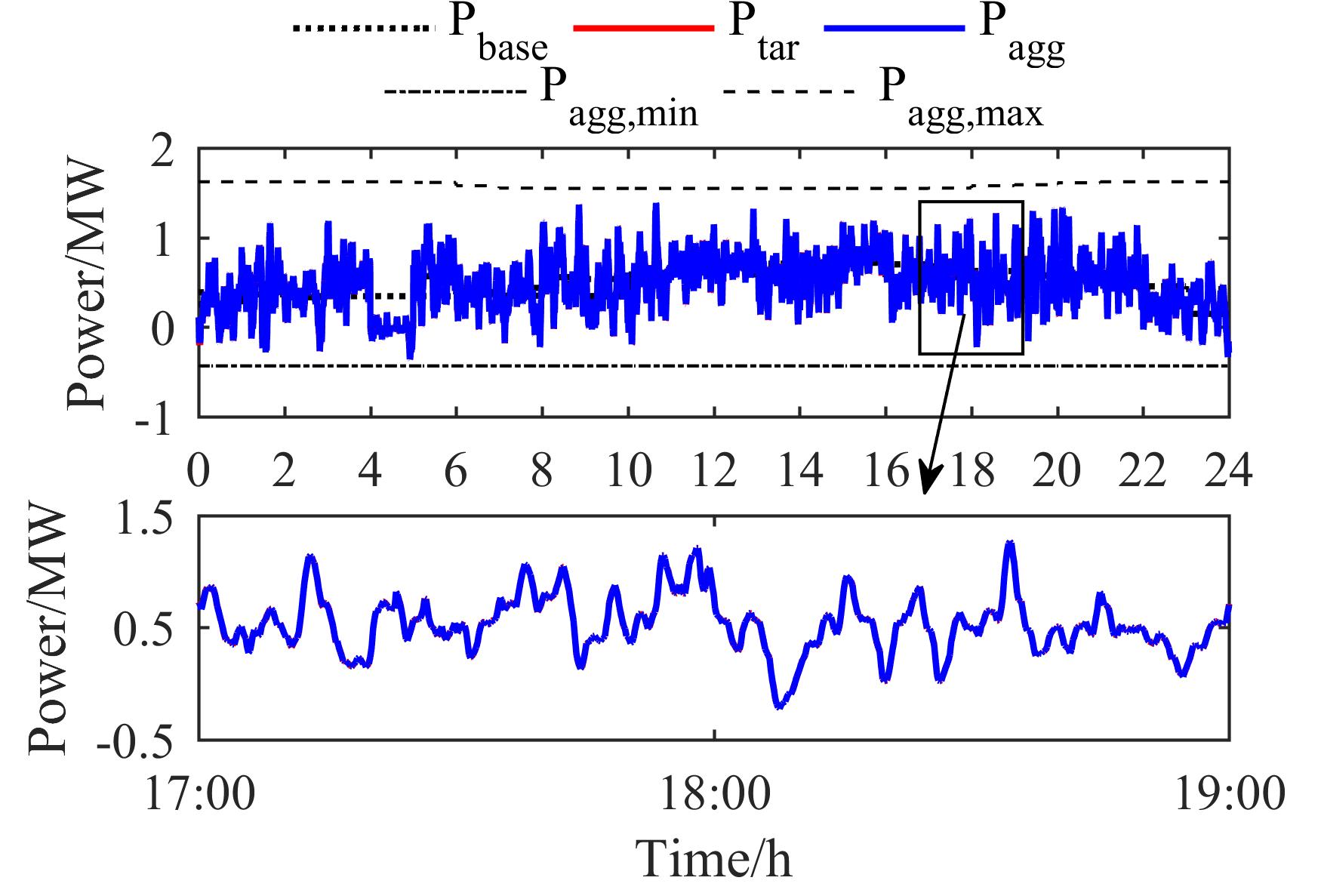}
\caption{Tracking performance in case 2.}
\label{fig:ER_tracking}
\end{figure}

 The target power $P_{tar}$ in this case is calculated by Eq.\eqref{eq:Ptar}, and the tracking performance is shown in Fig.\ref{fig:ER_tracking}. The hourly value of $\omega_{score}$ and $\omega_{mile}$ are shown in Fig.\ref{fig:score_mile}. According to the result, $\omega_{score}$ can basically reach 0.95 under the proposed control framework. In comparison, when responding to regD signal, $\omega_{score}$ of a hydroelectric generator can be 0.7$\sim$0.8, while that of an electric energy storage can be higher than 0.9 \cite{PJM_regD_score}. Therefore, the simulation results demonstrate that the GESs discussed in this paper are very promising alternatives to provide fast and accurate frequency regulation services.

\begin{figure}[H]
\centering
\includegraphics[width=8 cm]{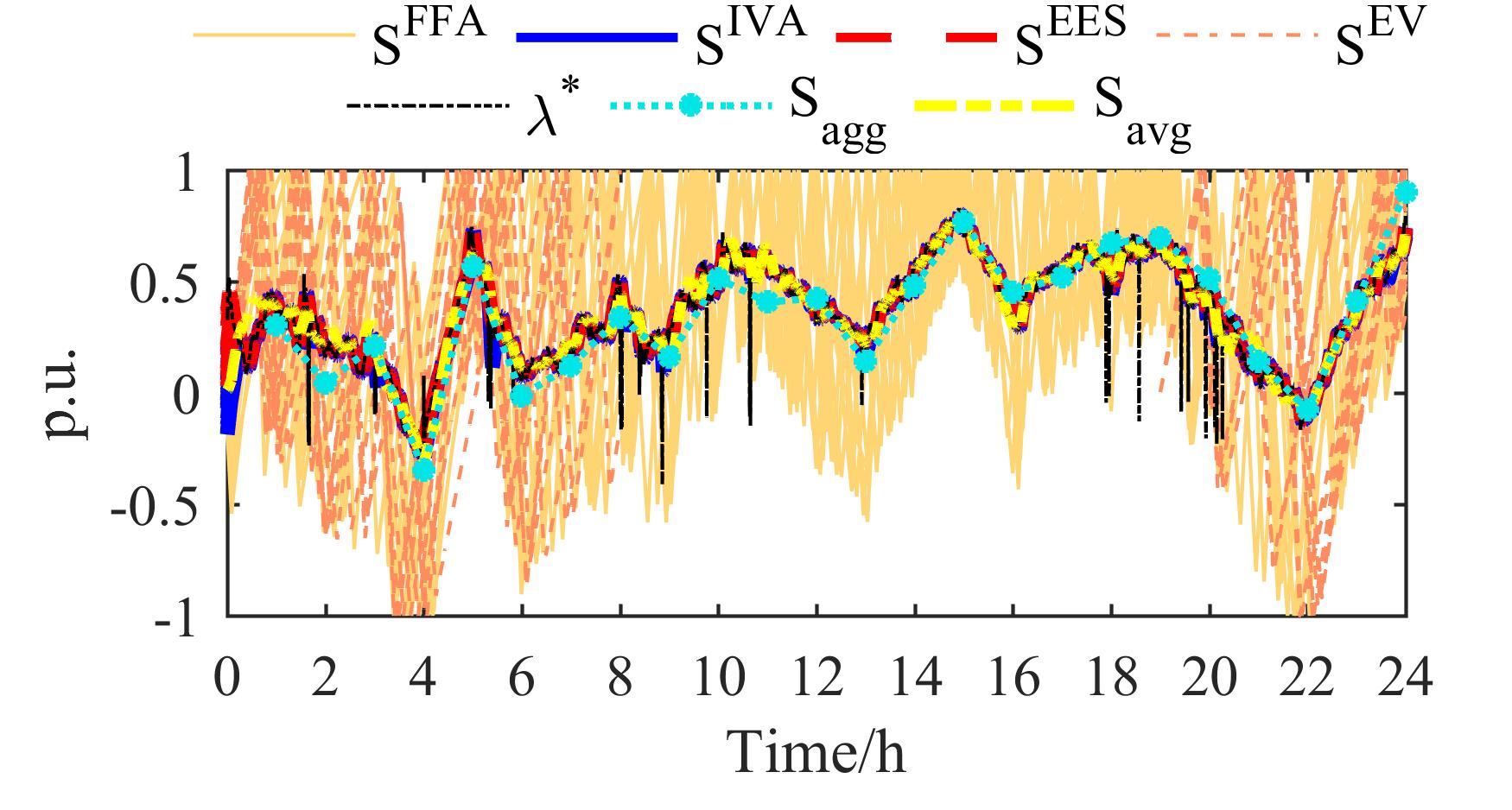}
\caption{DoS of different GESs and clearing price in case 2.}
\label{fig:ER_S}
\end{figure}

Trajectory of DoS and clearing price $\lambda^*$ are illustrated in Fig.\ref{fig:ER_S}. Compared with Fig.\ref{fig:EM_Savg}, since the GES cluster also needs to respond to the rapidly changing regD signal, some fluctuations and sudden changes can be observed in $\lambda^*$, which makes DoS of GESs unable to exactly follow $\lambda^*$. However, DoS always tends to approach $\lambda^*$, and thus the DoS-equality control is basically realized. In addition, at the end of some optimization cycles, e.g., at 10:00, 16:00, 24:00, the difference between $S_{avg}$ and $S_{agg}$ is a little larger than that in Fig.\ref{fig:EM_Savg}. It is because regD signal is not exactly zero-mean, which affects the actual energy consumption in each optimization cycle and enlarges the difference.

To analyse the economic benefits of the proposed method, the energy bill, regulation payments and total cost between the above cases (case 1 and case 2) and  the baseline case are compared. Hourly calculated cost is shown in Fig.\ref{fig:hourly_bill}, and Table.\ref{tab:cost_compare} lists the daily results. As can be observed, case 1 which only considers energy market can significantly reduce the energy bill compared with the baseline case. By optimally allocating flexility in two markets, case 2 further reduce the total cost compared to case 1, even though it receives a higher energy bill in the energy market. By providing fast regulation service, an LA obtains high payments from the regulation market, leading to a significant reduction in the total cost. Therefore, our method could achieve great economic benefits.

\begin{table}[H]
\caption{Comparison of costs (one day) in different cases.}
\centering
\begin{tabular}{cccc}
\toprule
 & \textbf{Baseline Case}	& \textbf{Case 1}	& \textbf{Case 2}\\
\midrule
Energy Bill/$\$$ & 1062.7 & 923.8 & 982.5\\
Change Rate	/\%	& / 	& -13.1			& -7.5\\
\midrule
Regulation Payments/$\$$	& 0	& 0	& 595.2\\
\midrule
Total Cost/$\$$	& 1062.7	& 923.8	& 387.3\\
Change Rate/\%	& /	& -13.1	&	-63.6\\
\bottomrule
\end{tabular}
\label{tab:cost_compare}
\end{table}
\subsection{Response Performance of Individual GESs}
\begin{figure}[H]
\centering
\includegraphics[width=16 cm]{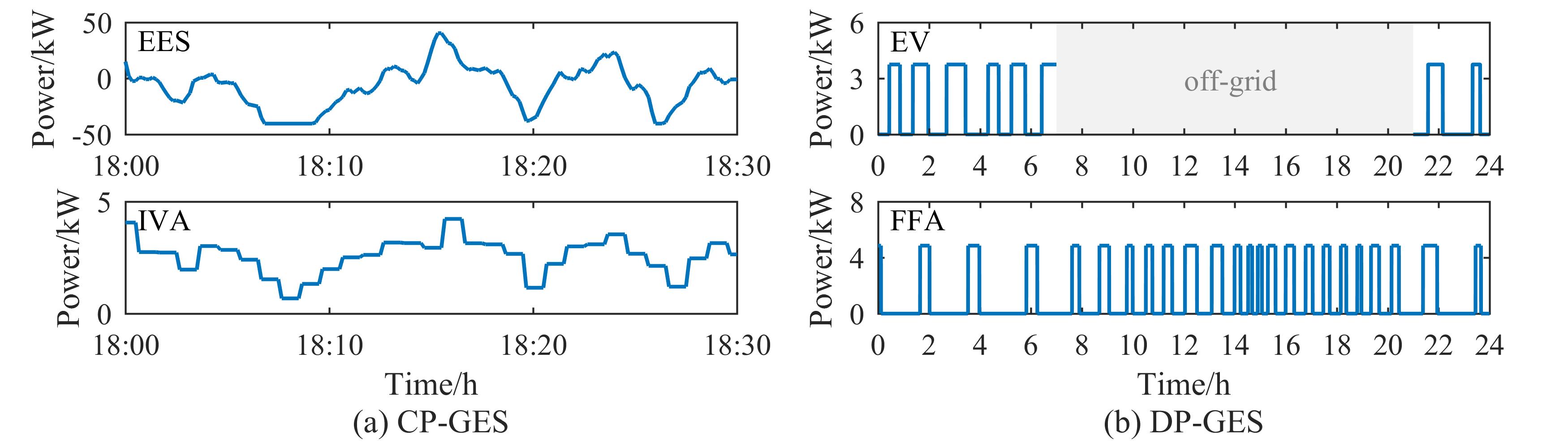}
\caption{Response power of a single GES in case 2.}
\label{fig:ER_single_load}
\end{figure}

The proposed control strategy ensures DoS-equality among GESs. While for different types of GESs, their response behaviour may be different due to their distinguishing features. To demonstrate this, pick one GES randomly from each type of GESs, and plot their response power in Fig.\ref{fig:ER_single_load}.

For CP-GESs, this paper allows different response cycles. For example, an EES adjusts its response power each 10s ($t_{res}=10s$), while an IVA adjusts its power every 1min ($t_{res}=60s$) considering its response ability, thus an IVA's response power is stair-shaped as shown in Fig.\ref{fig:ER_single_load}(a).

 DP-GESs adjust their response power by regulating the duty cycle. Among them, an EV's duty cycle at different time is basically similar, because its operation is irrelevant to external conditions and mainly depends on the user's charging pattern. In contrast, the required power of an FFA varies over time as it is significantly affected by environment conditions, e.g., outdoor temperature. For example, it can be observed from Fig.\ref{fig:ER_single_load}(b) that the duty cycle of an FFA during 13:00-16:00 is higher than that in the rest of the day as more cooling energy is required during these time slots.

\begin{figure}[H]
\centering
\includegraphics[width=8 cm]{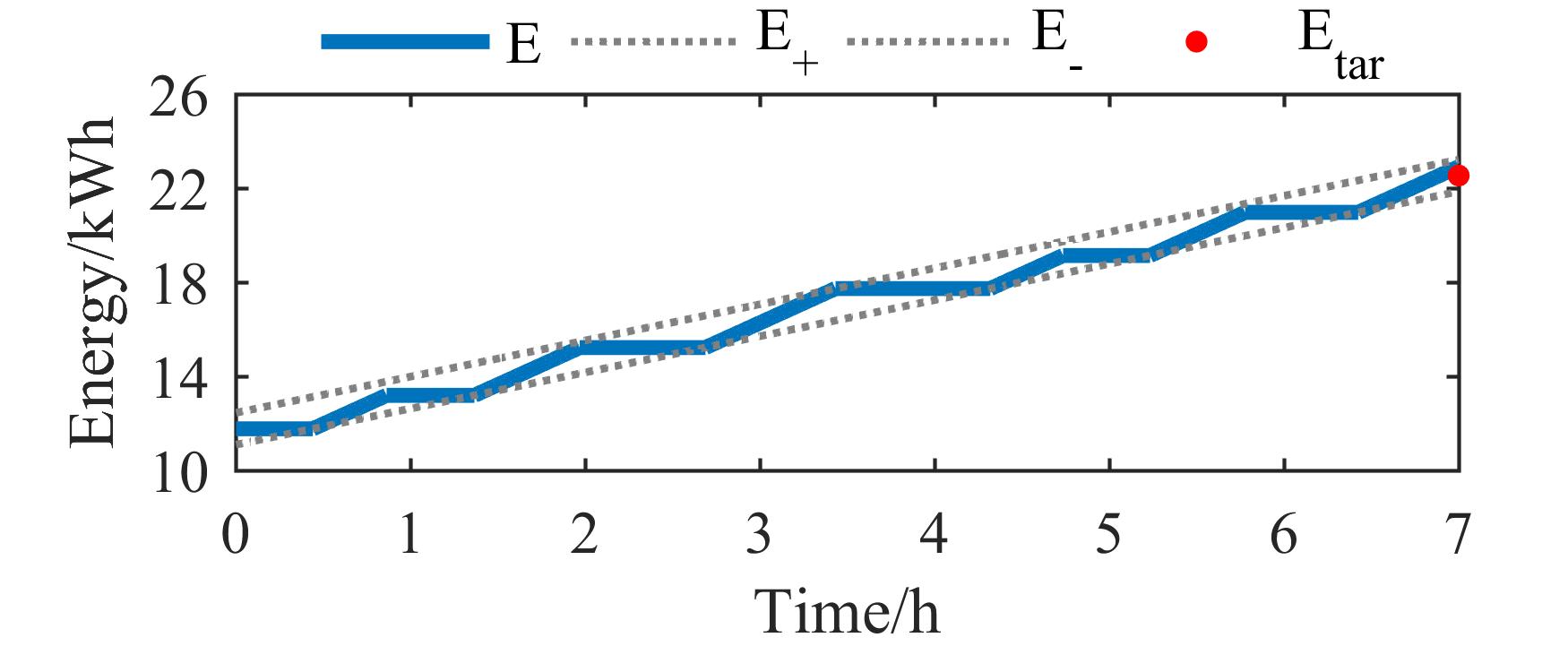}
\caption{Electric energy of a single EV in case 2.}
\label{fig:ER_single_ev}
\end{figure}

In addition, an EV should ensure that the electric energy reaches its target value $E_{tar}$ at departure time. The change of electric energy is shown in Fig.\ref{fig:ER_single_ev}, where $E_+=E^{exp}+r\%C^{EV}$, $E_-=E^{exp}-r\%C^{EV}$. As we can see, the hysteretic model and control strategy used in this paper can guarantee that the difference between the energy at departure time and the target energy $E_{tar}$ would not exceed $\pm r\%C^{EV}$.

\section{Conclusions}
\label{sec:conclusion}
In this paper, a unified coordination method is developed for large-scale heterogeneous GESs to participate in both energy and regulation market.

A generalized state variable referred to as DoS is first defined for GESs. The dynamic models with a unified form are then developed for different GESs. In real-time control, a market-based coordination framework is adopted, and a DoS-equality control method is then developed by construction of generalized demand curves for both GESs operating at continuous power and GESs with discrete states. Based on the unified dynamic models and the DoS-equality control feature, a low-dimensional aggregate dynamic model for a GES cluster is derived. At last, an optimization model aiming to allocate the flexibility of a GES cluster into both the energy market and the regulation market is developed, which uses the aggregate model to significantly reduce the mathematical complexity of the optimization problem.

The control framework has unified uplink/downlink information interfaces and supports a tree-like structure in both real-time coordination stage and optimization control stage, which makes it flexible, scalable and suitable for the control of large-scale GESs. Simulation results demonstrate that the aggregate model well describes the dynamic behaviour of a GES cluster. Additionally, the real-time control method can track the target power accurately while satisfying diversified requirements of different GESs and ensuring control fairness. It is also shown that an LA could gain considerable energy bill savings and high payments by participating in both energy and regulation markets.

However, simulations in this paper are based on ideal communication system and perfect model parameter identification. Future work would study the robustness of our method under communication problems and model errors.
\vspace{6pt}



\authorcontributions{Conceptualization and review: Peichao Zhang; simulation and writing: Yao Yao; review and editing: Sijie Chen.}

\funding{This research was funded by National Key R\&D Program of China (2018YFB0905000).}


\conflictsofinterest{The authors declare no conflict of interest}

\abbreviations{The following abbreviations are used in this manuscript:\\

\noindent
\begin{tabular}{@{}ll}
GES & Generalized Energy Storage\\
EES & Electric Energy Storage\\
EV & Electric Vehicle\\
FFA & Fixed-Frequency Air-conditioner\\
IVA	& Inverter Air-conditioner\\
TCL	& Thermostatically Controlled Load\\
DoS	& Degree of Satisfaction\\
LA	& Load Aggregator
\end{tabular}}

\appendixtitles{no} 
\appendixsections{multiple} 
\appendix
\section{Derivation of the function $g^{IVA}_P(T_{tar},t_p)$}
\label{sec:appendix_gP}
The analytical solution of Eq.\ref{eq:IVA_thermal} can be described as:
\begin{equation}
T_{a,i}(t)=(T_{a,i}(0)-T_{o,i}(0)+\frac{1}{a_iC_{th,i}}Q^{IVA}_{i,k})e^{-a_i t}+T_{o,i}(0)-\frac{1}{a_iC_{th,i}}Q^{IVA}_{i,k}
\end{equation}
where $T_{a,i}(0)$ and $T_{o,i}(0)$ denote the current indoor air temperature and outdoor temperature of IVA $i$; $T_{a,i}(t)$ denotes the indoor air temperature at time $t$.

To derive the electric power required to make $T_{a,i}$ change from $T_{a,i}(0)$ to $T_{a,i}(t)$ over a certain period of time $t_p$, we let $t=t_d$, $T_{a,i}(t_p)=T_{tar}$, and the required heat rate $\hat{Q}^{IVA}_i$ can be calculated by
\begin{equation}
\hat{Q}^{IVA}_i=\frac{(T_{tar}-T_{o,i}(0))-(T_{a,i}(0)-T_{o,i}(0))e^{-a_i t_p}}{R_{th,i}(e^{-a_i t_p}-1)}
\end{equation}
According to Eq.\eqref{eq:IVA_linear}, the required electric power $\hat{P}^{IVA}_i$ can be derived by
\begin{equation}
\hat{P}^{IVA}_i=g^{IVA}_P(T_{tar},t_p)=p_1\frac{\hat{Q}^{IVA}_i-q_2}{q_1}+p_2
\end{equation}

\section{}
\unskip
\begin{figure}[H]
\centering
\includegraphics[width=8 cm]{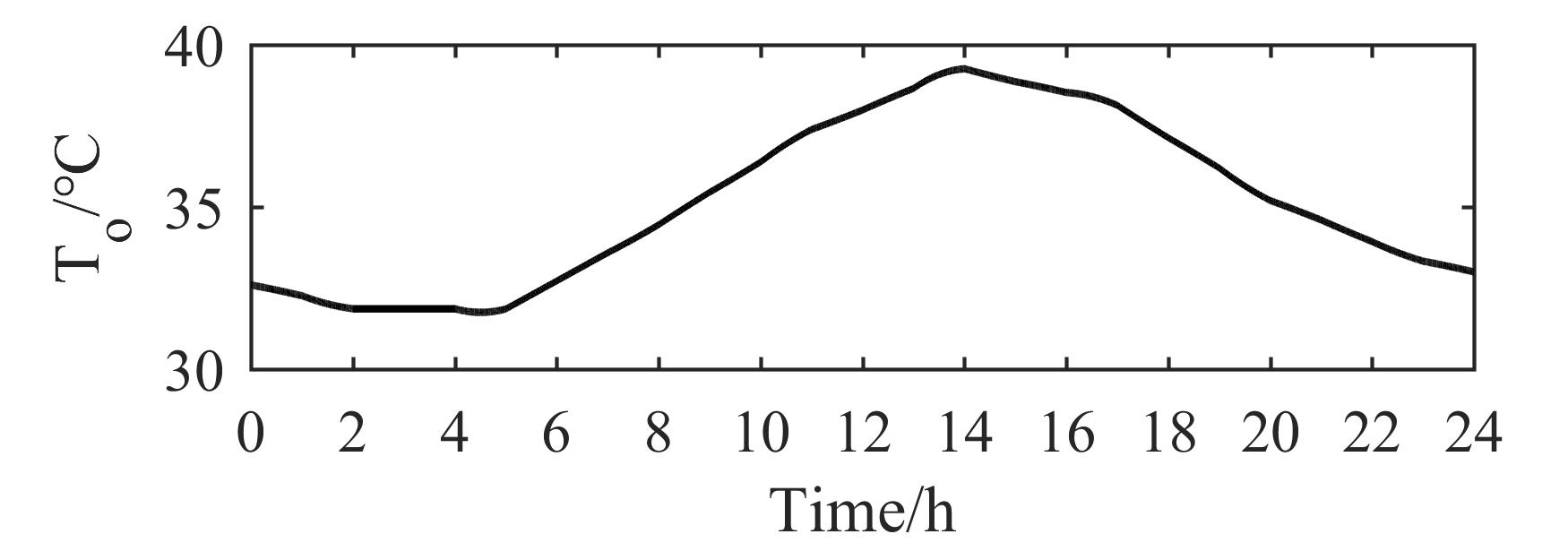}
\caption{Outdoor temperature.}
\label{fig:To}
\end{figure}

\begin{figure}[H]
\centering
\includegraphics[width=8 cm]{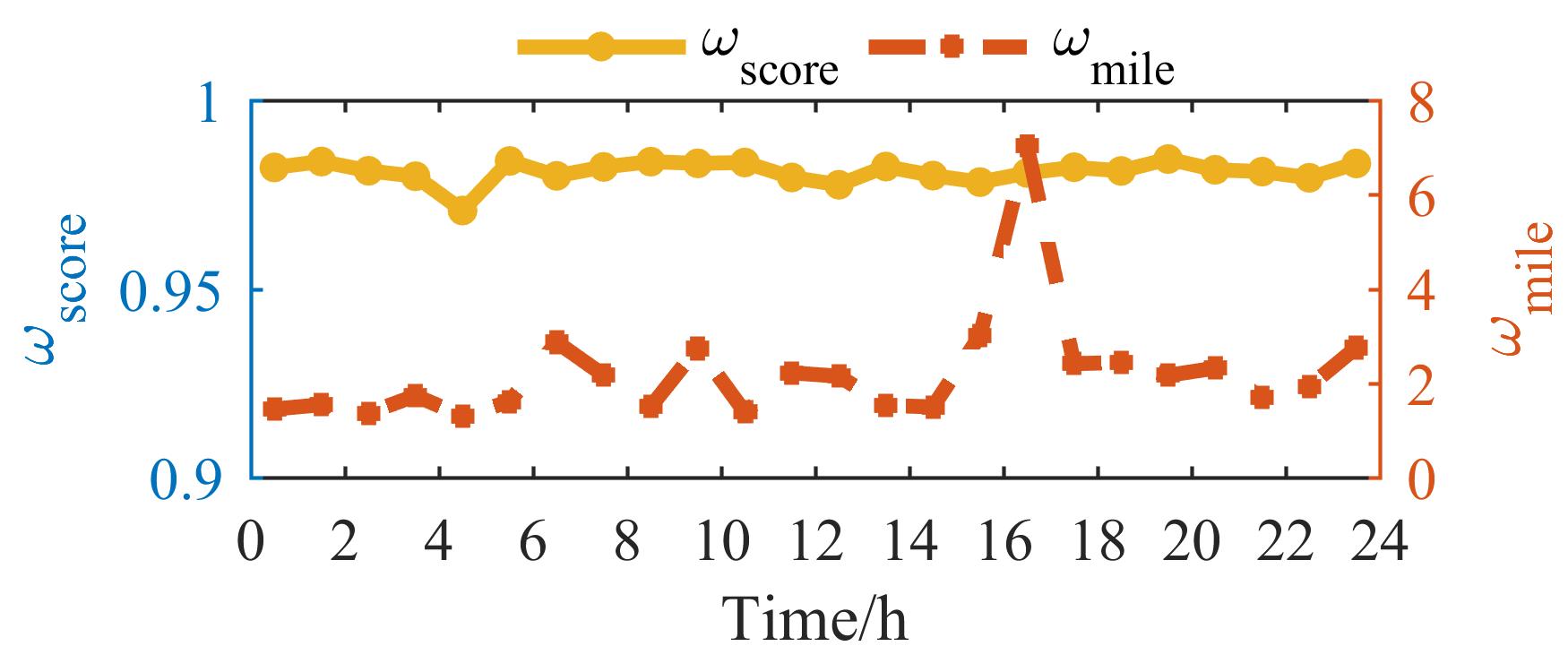}
\caption{$\omega_{score}$ and $\omega_{mile}$.}
\label{fig:score_mile}
\end{figure}

\begin{figure}[H]
\centering
\includegraphics[width=8 cm]{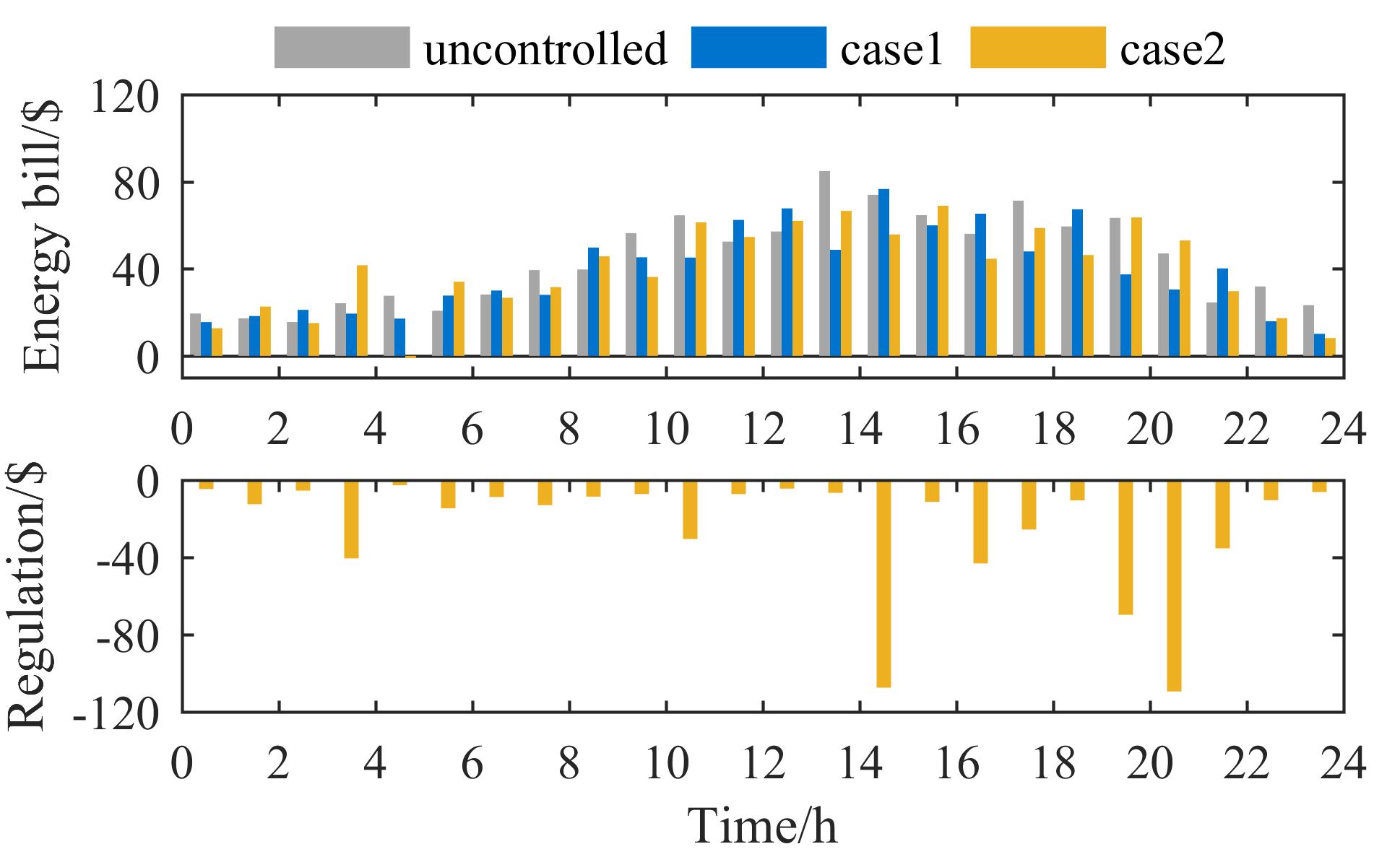}
\caption{Hourly energy bill and regulation payments.}
\label{fig:hourly_bill}
\end{figure}


\reftitle{References}


\externalbibliography{yes}
\bibliography{mdpi}
\end{document}